\documentstyle[11pt]{article}

\def\boxit#1{
\vbox{\hrule height0.5pt\hbox{\vrule width0.5pt\kern10pt\vbox{
\kern10pt#1\kern10pt}
\kern10pt\vrule width0.5pt}\hrule height0.5pt}}
\renewcommand{\theequation}{\arabic{section}.\arabic{equation}}
\def\bild#1\over#2{\mathrel{\mathop{\kern5pt #1}\limits_{#2}}}

\title{{\huge Quantum Chaotic Dynamics and Random Polynomials}}
\author{}
\date{{\LARGE E. Bogomolny, O. Bohigas and P. Leb{\oe}uf} \\Division
de Physique Th\'eorique\footnote{Unit\'e de recherche des Universit\'es de
Paris XI et Paris VI associ\'ee au CNRS}\\ Institut de Physique
Nucl\'eaire\\91406 Orsay Cedex, France \vspace{0.15in} \\}

\begin{document}
\baselineskip 0.25in
\setcounter{page}{0}
\maketitle

\vspace{0.4in}

\begin{abstract}
We investigate the distribution of roots of polynomials of high degree with
random coefficients which, among others, appear naturally in the context of
"quantum chaotic dynamics". It is shown that under quite general conditions
their roots tend to concentrate near the unit circle in the complex plane. In
order to further increase this tendency, we study in detail the particular case
of self-inversive random polynomials and show that for them a finite portion of
all roots lies exactly on the unit circle. Correlation functions of these roots
are also computed analytically, and compared to the correlations of eigenvalues
of random matrices. The problem of ergodicity of chaotic wave-functions is also
considered. For that purpose we introduce a family of random polynomials whose
roots spread uniformly over phase space. While these results are consistent
with random matrix theory predictions, they provide a new and different insight
into the problem of quantum ergodicity. Special attention is devoted all over
the paper to the role of symmetries in the distribution of roots of random
polynomials. 
\end{abstract}
\vspace{0.4in}


\section{Introduction}

Semiclassical approximations for multidimensional quantum systems and the
manifestations of chaotic behaviour in quantum mechanics have attracted wide
attention during the last years (see e.g. \cite{lesh},~\cite{coph} and
references therein). In these approximations one quite often needs to locate
the roots of polynomials of high degree whose coefficients are rapidly-varying
erratic functions of the energy. As a consequence, these coefficients may be
considered as random variables, even in a small energy interval. Though the
distribution of roots of polynomials with random coefficients have been studied
in the past (see e.g. \cite{rp}-\cite{kac}), its relevance with respect to the
specific problems which naturally arise in the context of quantum chaotic
dynamics (and in other domains of physics as well) has been underestimated, and
a number of elementary and basic questions have not yet been solved. 

The main purpose of this paper is to study properties of random polynomials
with emphasis on the above mentioned connection. A short version containing
some of our results was already published in \cite{bbl}. Except for some
general symmetry considerations, the coefficients of the polynomials will be
considered independent random variables. This assumption is in some cases
justified with physical arguments, and in others just made because of
mathematical simplicity. Our investigations could have direct application in 
other fields. In fact, zeros in the complex plane of polynomials with random
coefficients occur in a variety of problems of science and engineering (zeros
of the partition function in statistical mechanics, theory of noise, etc). The
present investigation may also be of interest if one views zeros of polynomials
as interacting particles in two dimensions, as, for instance, eigenvalues of
random asymmetric matrices can be physically interpreted as a two-dimensional
electron gas confined in a disk \cite{gin,jan}. 

The use of a statistical approach in the description of complex systems is an
old idea. In particular the random matrix theory (RMT), originally formulated
in the context of nuclear physics, has had a great success and impact in the
study of quantum chaotic dynamics\footnote{we will use this expression to avoid
the more proper but lengthy term 'quantum mechanics of a classically chaotic
system'.}  and disordered systems \cite{lesh,alw}. As it will become clearer in
the following sections, the statistical analysis of these systems by random
polynomials is, in some sense, complementary to the RMT. 

In Section 2 we consider general random polynomials of degree $N$ whose
coefficients are independent random variables having zero mean. We show that
under quite general conditions their roots tend to concentrate in an annulus
near the unit circle of the complex plane, and that the width of this annulus
goes to zero as $N\rightarrow \infty$. In \cite{rp} this result was proved by a
different method for the particular case when all second moments of the
coefficients are equal. Our method, based on the existence of a saddle point
configuration, seems to be more general and physically transparent. 

Aside from section 2, which constitutes a general introduction, the paper can
be divided into two main parts (sections 3 and 4) which are to a large extent
independent. In Section 3 we investigate the important case of self--inversive
polynomials, namely polynomials whose coefficients satisfy Eq.(\ref{2e6b}) 
below and whose roots are hence distributed symmetrically with respect to the
unit circle. This type of polynomials appear when considering either the
semiclassical quantization via the transfer-operator method \cite{bogo} or
quantum maps \cite{bbtv,sv}. We prove that for them a finite fraction of the
roots lies exactly on the unit circle. Moreover, we compute analytically the
two--point correlation function of these roots. We find that there is linear
repulsion between them at short distances, and compare to the Gaussian
Orthogonal Ensemble of RMT \cite{mehta,boh}. The possibility of locating in a
statistical sense all the roots on the unit circle is also discussed. 

In Section 4 we consider the properties of eigenfunctions of chaotic systems
and their relation with the classical concept of ergodicity, in the spirit of
Refs \cite{lv,leb1}. Using a phase-space representation, we write down the
eigenfunctions of spin systems in a polynomial form and consider the
coefficients to be random. We then analyse how the roots of these polynomials
are distributed in phase space. The second moments of the coefficients now
grow with $N$ so fast that the resulting distribution of roots turns out to
be uniform over the phase space, a sort of quantum ergodicity. We moreover
compare this result with the predictions concerning the coefficients of the
eigenfunctions of a random matrix ensemble. 

It turns out that the existence of a symmetry  of the roots with respect to a
line increases considerably the probability of finding roots exactly on that
line. In subsection 4.2 we consider the influence of symmetries on the
distribution of the roots of chaotic eigenstates. For that purpose
we analyze the eigenfunctions of a certain quantum map having two antiunitary
symmetries and show numerically that their roots tend to concentrate over the
associated phase-space symmetry lines. We also investigate how this phenomenon 
disappears as the symmetries are broken.

In the appendices we explain in detail our computations.

\section{Some General Properties of Random Polynomials}

Given a distribution function
\begin{equation} \label{1e2}
\int {\cal D} (a_0, a_1, \ldots, a_N) d^2 a_0 d^2a_1 \ldots d^2 a_N
\end{equation} 
for some complex coefficients $\left\{ a_k\right\}$, we are interested in the
distribution in the complex plane of the roots $\left\{ z_k\right\}$, $k= 1,
\ldots , N$ of the polynomial 
\begin{equation} \label{1e1}
P(z)= a_0+a_1 z+a_2 z^2+\ldots + a_N z^N . 
\end{equation}
For clarity, we consider the particular case where the real and imaginary part
of the coefficients $\left\{ a_k\right\}$ are independent normally distributed
real-valued random variables with zero mean and standard deviation $\sigma_k$
(henceforth denoted as a GRI distribution) 
\begin{equation} \label{1e4}
{\cal D} (a_0, \ldots , a_N) = 
 \prod\limits^N_{k=0} 
\frac{1}{2\pi \sigma^2_k} \exp \left( -\frac{|a_k|^2}{2\sigma^2_k}\right) \ . 
\end{equation}
The joint probability density for the zeros is obtained by changing from the
variables $\left\{ a_0, a_1, a_2, \ldots , a_N\right\}$ to the variables
$\left\{ a_N, z_1, z_2, \ldots , z_N\right\}$ using the standard formulae 
$$ \begin{array}{l} 
a_{N-k}=\left(-1\right)^k a_N \, u_k (z), \;\;\;\;\;\;\;\;\;\;\; k=1, \cdots ,
N\\ 
u_k (z)=\sum\limits_{1=i_1<i_2<\ldots <i_k=N} z_{i_1}z_{i_2}\cdots z_{i_k},
\end{array}
$$
and integrating over $a_N$, with the result
\begin{equation} \label{1e5} 
{\cal D}(z_1, \ldots , z_N) = C_N \frac{D_N (z)}{\left[ G_N(z)\right]^{N+1}},
\end{equation}
where

\begin{equation} \label{1e5a} \begin{array}{l}
C_N=\frac{N!}{\pi^N} \prod\limits^N_{k=0} \eta_k,\;\;\;\;\;
\eta_k =\left(\frac{\sigma_N}{\sigma_{N-k}}\right)^2,\\
G_N(z)=1+\sum\limits_{k=1}^{N} \eta_k \left| u_k(z) \right|^2,\\
D_N (z)=\prod\limits_{j<k} |z_j- z_k|^2.
\end{array} 
\end{equation}
The factor $D_N(z)$ comes from the Jacobian of the transformation. 

Some properties of this distribution are:
\begin{itemize} 
\item[(i)] if for all $k$  $\sigma_k =\sigma$ ,  then ${\cal D} (z_1, \ldots ,
z_N)$ is independent of $\sigma$. This is a consequence of the fact that the
roots of $P(z)$ are unchanged by multiplying $P(z)$ by a constant. 

\item[(ii)] ${\cal D}(z_1, \ldots , z_N)$ is invariant under a rotation of 
the coordinates in the complex plane, i.e. 
$$ 
{\cal D}(z_1, 
\ldots , z_N)= {\cal D}(z_1 e^{i \varphi} , \ldots , z_N e^{i \varphi})\ .
$$

\item[(iii)] if $\sigma_{N-k} =\sigma_k$, then ${\cal D}(z_1, \ldots , z_N)$ is
invariant under inversion with respect to the unit circle ${\cal C}$ in the
complex plane, 
$${\cal D}(z_1, \ldots , z_N)= {\cal D}(1/{\bar z}_1, \ldots , 1/{\bar z}_N),$$
\end{itemize}
where the bar means complex conjugate.

In order to find the most probable distribution of roots, we have to locate the
maximum of ${\cal D}$. The equation for the extrema $d \log {\cal D} /d{\bar z}
_p=0$ can be transformed into the following form: 
\begin{equation} \label{1e5b}
\sum_{k=1}^{N}\eta_{k}u_k  
\left [ \overline{\left (\frac{du_k}{dz_p}\right )}- g_p{\bar u}_k \right ]=g_p,
\;\;\;\;\;\;\;\;\;\;\;\;\;\;\; p=1,\ldots,N
\end{equation}
where
$$
g_p=\frac{1}{(N+1)} \sum\limits_{j\neq p}\frac{1}{{\bar z} _p-{\bar z} _j} \ .
$$
One can easily check that the following configuration is a solution of these
equations:
\begin{equation} \label{1e3}
z_k = r_{N} \exp \left[ {\rm i} \frac{2\pi}{N} k + {\rm i} \phi \right], \quad
k = 1, \ldots , N \ , \end{equation} 
where
$$
r_{N} =\left(\frac{N-1}{N+3}\right)^{1/2N}
\left(\frac{\sigma_0}{\sigma_{N}}\right)^{1/N} \buildrel N \to \infty
\over{\simeq} 1 + \frac{1}{N} \ln \left(\frac{\sigma_0}{\sigma_N}\right) \ . 
$$
We will refer to this configuration as to the crystal solution. To prove that
it obeys Eq.(\ref{1e5b}) we remark that the $z_k$ given by Eq.(\ref{1e3}) are
solutions of the equation $z^N=const$, and therefore for this solution all
$u_k$ with $k=1,2,\ldots,N-1$ vanish and Eq.(\ref{1e5b}) reduces to 
\begin{equation} \label{1e6}
\eta_{N}u_N \left [ \overline{\left (\frac{du_N}{dz_p}\right )}- g_p{\bar u}_N
\right ]=g_p \;\;\;\;\;\;\;\;\;\;\;\;\;\;\; p=1,\ldots,N.
\end{equation}
But $du_N/dz_p=u_N/z_p$ and $g_p=C/{\bar z}_p$ where 
$$
C=\frac{1}{N+1} \sum_{j=1}^{N-1} \frac{1}{1-\exp ({\rm i} 2\pi j/N)}=
\frac{1}{2}\frac{(N-1)}{(N+1)}. 
$$
Then Eq.(\ref{1e6}) becomes
$$\eta_N |u_N|^2 (1-C) = C \ ; $$
using the definitions of $\eta_N$, $u_N$ and $C$, the reader can verify that
this equation is satisfied. Therefore all Eqs.(\ref{1e5b}) for $p=1,\ldots,N$
will be fulfilled by the crystal solution (\ref{1e3}). 

It follows that (except in the case of an exponential (or faster) dependence of
$\sigma_0/ \sigma_N$ on $N$) in the limit $N \to \infty$ the radius $r_N$ tends
always to one. Moreover, the phase $\phi$ in Eq.(\ref{1e3}) is arbitrary
because of the property (ii). In the particular case $\sigma_0 = \sigma_N$,
$r_N$ is equal to one, as it must be according to property (iii). 

The existence of the crystal solution implies that the distribution of roots
${\cal D} (z_1, \ldots , z_N)$ will always have an extremum on the line $\left|
z\right|=r_N$. But the actual distribution of roots in the complex plane
depends, however, on the second derivatives of $\log {\cal D}$. There is a
direct connection between their magnitude and the dominance of the crystal
solution. For example, putting $\sigma_0 =\sigma_N$ and choosing all the other
$\sigma$'s such that $N \eta_k \rightarrow 0, \, (k=1, \ldots , N-1)$ as $N \to
\infty$, one concludes  that  the crystal solution will not dominate the
distribution since the intermediate terms in the denominator of (\ref{1e5})
automatically cancel, independently of the crystal solution. An extreme case of
this behaviour is what we call, for reasons which will become clear later, the
$SU(2)$ polynomials, for which $\sigma_k/ \sigma_N=\sqrt{C^k_N}$ where $C^k_N$ 
are Newton's binomial coefficients. As we shall see in section 4, for these
polynomials the roots spread all over the complex plane. 

On the contrary, the case  $N\eta_k \rightarrow \infty$  enforces the
crystal solution since the factors $u_k$ in the denominator of (\ref{1e5}) 
are multiplied by an increasing function of $N$. We shall consider in more 
detail several cases of this type in the next sections. 

These simple considerations show that under  quite general conditions the roots
of random polynomials of high degree for which the mean value of the
coefficients is equal to zero will tend to concentrate near the circle $\left|
z\right|=1$. In order to illustrate this with a numerical example, in Fig.1a
 are plotted, in the complex plane, the roots of 200 trials of a
polynomial of degree $N=48$ whose coefficients are GRI-distributed all having
the same second moment. The observed distribution clearly satisfies the
properties (ii) and (iii). 

On the other hand, it can be easily proved that if the mean value of the
coefficients are non-zero, the roots tend to concentrate around the roots of
the mean polynomial \cite{rp}. 

\section{The Self-Inverse Symmetry} 
\setcounter{equation}{0}

As shown in the previous section and as illustrated in Fig.1a, in the large-$N$
limit and under certain circumstances the roots of random polynomials tend to
concentrate around the unit circle but not, in general, on it. We now want to
study some simple  conditions to be imposed over the coefficients of a random
polynomial in order to locate as much zeros as possible {\sl exactly} on the
unit circle. Among other areas of physics and mathematics, this problem is of
interest in the context of semiclassical approximations in quantum  mechanics
because of the following reasons. 

Some recent methods to solve quantum problems incorporate the physical
information into a transfer operator $T$ which is an $N \times N$ unitary
matrix \cite{bogo}. The operator $T$ has $N$ complex
eigenvalues $\left\{ z_k\right\}, \, k=1, \ldots ,N$ lying on the unit circle
$\cal C$ of the complex plane, which are determined by the roots of the
characteristic polynomial 
\begin{equation} \label{2e1}
P(z)= \det (z - T) = \sum^N_{k=0} a_k \, z^k \ . 
\end{equation}
The eigenvalues are functions of the energy of the system, $z_k=z_k(E)$, since
$T$ is a function of $E$. They move on $\cal C$ as $E$ is varied, and the
quantization condition, which takes the form of a  Fredholm determinant, states
that whenever one of the roots crosses the point $z=1$ then the corresponding
energy is an eigenvalue of the system
\begin{equation} \label{2e2}
\det \left(1- T(E)\right) = 0 \ . 
\end{equation}
A similar equation but with no energy dependence is obtained when considering 
the quantization of maps, where $T$ is now the one step unitary evolution
operator of the map \cite{bbtv}. Since $T$ is a finite matrix, it has $N$ nontrivial
invariants which can be chosen either as the coefficients $\left\{ a_k\right\}$
of the characteristic polynomial, or as the traces of the first $N$ powers of
$T$ 
\begin{equation} \label{2e3} M_L= {\rm Tr} \ T^L (E) = \sum^N_{k=1} z^L_k \ , 
\;\;\;\;\;\;\;\;\;\; L=0, \ldots ,N \ .
\end{equation}
One can express one set of invariants in terms of the other set by means of the 
following recurrent formulas
\begin{equation} \label{2e4}
a_{N-k}= -\frac{1}{k} \sum^k_{j=1} a_{N-k+j} \, M_j \ , \;\;\;\;\;\;\;\;
k= 1, \ldots , N \ , 
\end{equation}
with $a_N = 1$ by definition of $P (z)$. These relations are of interest
because in the semiclassical limit $N \to \infty$ one can express the traces of
the powers of $T$ as an amplitude sum over the classical periodic orbits 
\begin{equation} \label{2e6}
M_L \simeq \sum_{\gamma(L)} A_\gamma (E)  
e^{{\rm i} S_\gamma (E) /\hbar -{\rm i} \pi \mu_\gamma /2}  \ .
\end{equation}
Here, $\gamma(L)$ are all the periodic orbits corresponding to $L$ iterations of
the initial map at energy $E$ (including repetitions), $S_\gamma(E)$ is the
action of the periodic orbit, $\mu_\gamma$ is the Maslov index and
$A_{\gamma}(E)$  is a real function depending on the stability of the orbit.
Inserting (\ref{2e6}) into (\ref{2e4}) we obtain a semiclassical approximation
for each coefficient $a_k$, now written in terms of sums over all the periodic
orbits up to period $k$ (the special combinations between them given by
Eqs.(\ref{2e4}) are called pseudo-orbits). To compute all the coefficients, all
the periodic orbits up to period $N$ are needed. 

Since the action $S_\gamma (E)$ is a function of the energy,  in the
semiclassical limit the moments are rapidly varying functions of the energy 
and the coefficients $a_k$ are sums of products of these rapidly varying 
functions. It thus seems natural to adopt a statistical approach and to consider
these coefficients as random variables. This way of proceeding differs from the
usual statistical approach to complex systems in which, instead of the
coefficients of its characteristic polynomial, the matrix elements of a
relevant operator are assumed to be random. 

There is, however, an intrinsic difficulty when approximating the coefficients 
of the characteristic polynomial by Eqs.(\ref{2e4}) and (\ref{2e6}), or more 
generally  when considering them as independent variables. We are ignoring the
correlations existing among them that guarantee the unitarity of $T$. (Without
correlations and as was pointed out in section 2, the roots will lie close to
$\cal C$ but not on it.)  A consequence of this is that the eigenvalues of
$T$ are no more located on ${\cal C}$, and Eq.(\ref{2e2}) fails to determine the
full spectrum (typically, some of the eigenvalues are missed). 

It may therefore be useful to find some simple conditions to be imposed on the
coefficients $\left\{ a_k\right\}$ in order to restore -- at least partially --
the unitarity of $T$. A necessary but not sufficient condition is the
self-inversive (SI) property 
\begin{equation} \label{2e6b}
a_{N-k} =\exp\left({\rm i}\Theta\right) {\bar a}_k 
\end{equation}
which can easily be obtained from Eq.(\ref{2e1}) factorizing the polynomial and
making the substitution $z_k = \exp \left({\rm i} \theta_k\right)$. $\Theta$ is
a real function of the energy, $\Theta=\pi N +\sum\limits^N_{k=1} \theta_k
(E)$. In the semiclassical limit \cite{bogo} 
$$\Theta \simeq \pi \overline{N(E)},$$
where $\overline{N(E)}$ is the mean number of levels with energy less than $E$.
Polynomials obeying Eq.(\ref{2e6b}) satisfy a functional equation 
\begin{equation} \label{2e6c}
\overline{
P(1/\bar z) }= {\rm e}^{-{\rm i} \Theta} P(z)/z^N \ . 
\end{equation}
It follows from this equation that the symmetry (\ref{2e6b}) of the
coefficients is reflected into a symmetry of the zeros: if $z_k $ is a root,
then $1/\bar z_k$ is also a root, i.e. the roots either lie on $\cal C$ or are
symmetrically located under inversion with respect to it. 

From the semiclassical point of view, the advantages of imposing the
self-inversive symmetry are twofold: firstly because it implements, in a simple
manner, part of the unitarity of $T$; and secondly because it lowers the number
of periodic orbits to be computed. In fact, we only need to know the periodic
orbits up to the period $N/2$ instead of $N$, since we only need to compute
half of the coefficients (the others being determined by symmetry) (see e.g.
\cite{bs,sv}). 

\subsection{The fraction of roots lying on ${\cal C}$}

Being a necessary but not sufficient condition we don't know, however, how many
zeros are located on the unit circle by the SI symmetry. In order to answer this
question, we compute the fraction of zeros lying on ${\cal C}$ for SI
polynomials of the form (remember that $a_N=1$) 
\begin{equation}  \label{2e7}
P (z)=1+ \sum^{N-1}_{k=1} a_k z^k + z^N \ , \;\;\;\;\;\;\;\;\;\;\; 
a_{N-k}= {\bar a}_k \ ,
\end{equation} 
and consider the coefficients $\left\{ a_k \right\}, k=1, \ldots , (N-1)/2$ as
GRI distributed complex variables with arbitrary variances $\sigma^2_k$. For
simplicity we consider the particular case $\Theta = 0$ and consider $N$ to be
an odd integer, without loss of generality. By substitution $z= \exp ({\rm i}
\theta)$ in (\ref{2e7}), SI polynomials transform into real trigonometric
polynomials: 
\begin{equation} \label{2e9} 
f(\theta) = \frac{1}{2}{\rm e}^{-{\rm i} N \theta/2} P \left( {\rm e}^{{\rm i}
\theta}\right)= \cos \left(\frac{N}{2} \theta\right) + \sum^M_{k=1} \left\{ 
c_k \cos \left[ \left( \frac{N}{2} -k \right) \theta \right] + d_k \sin \left[
\left(\frac{N}{2} -k \right) \theta \right] \right\} \ , 
\end{equation}
where $M=(N-1)/2$, $c_k= \Re e (a_k)$ and $d_k=\Im m (a_k)$. The zeros of 
$P(z)$ lying on ${\cal C}$ correspond now to the real zeros of the real
function $f(\theta)$. 

The average fraction of zeros lying on ${\cal C}$ is defined as 
\begin{equation} \label{2e9b}
<\nu> =\frac{1}{N} \int^{2 \pi}_0 <\rho (\theta) > d\theta
\end{equation}
where $<\rho (\theta)>$ is the average density of zeros on ${\cal C}$
\begin{equation} \label{2e10} 
\rho (\theta)= \sum_k \delta (\theta-\theta_k)= \delta [f(\theta)]|f'
(\theta)| \ .
\end{equation} 
(Primes indicate derivative with respect to $\theta$.) To compute $<\rho
(\theta)>$, we use the method of Kac \cite{kac} which exploits the following
representations of $\delta[f]$ and $|f'|$
\begin{equation} \label{2e11}
\delta[f] =\int^\infty_{-\infty} \frac{d \xi}{2 \pi} {\rm e}^{{\rm i} \xi f} \
, \quad |f'|= \int^\infty_{- \infty} \frac{d \eta}{\pi \eta ^2} \left(1-{\rm
e}^{{\rm i}\eta f'} \right) 
 \ . \end{equation} 
The advantage of such representations is that, when performing the ensemble
average over the  coefficients 
\begin{equation} \label{2e12}
< \rho(\theta)>=\int {\cal D} (a_0,a_1, \ldots , a_N) \, \rho (\theta) \, 
d^2a_0 \ldots d^2 a_N 
\end{equation} 
the exponentiation of $f$ and $f'$ (who are linear function of the $\left\{ 
a_k\right\}$) allows an easy computation of the integrals.

The computation of $<\rho(\theta)>$ and $<\nu>$ for arbitrary $N$ and arbitrary
second moments $\left\{ \sigma_k \right\}$ are straightforward but lengthy, and
we include them in the appendix A. Here we give the result for the particular
case of constant second moments $\sigma_k= \sigma \, \forall k $ and in the
limit $N \rightarrow \infty$ (see Eqs.(\ref{2e18}),(\ref{2e31a}) below for the
exact answer for arbitrary $N$ and arbitrary variances in the case of a SI
polynomial of the form (\ref{2e15b})). We find that, to leading order in
$1/N$, the average density of roots depends on $\sigma$ through the scaled
parameter $\varepsilon = \sigma \sqrt{N}$, with the result 
\begin{equation} \label{2e13a}
<\rho (\theta)>\simeq \frac{N}{2} \exp \left[ -\cos^2\left(\frac{N}{2}
\theta\right)/ \varepsilon^2 \right]
\left\{\frac{1}{\sqrt{\pi}\varepsilon}\left| \sin \left(\frac{N}{2}
\theta\right)\right| + \frac{1}{\pi \sqrt{3}} \int^1_0 dx \, \exp \left[-3
\sin^2 \left(\frac{N}{2} \theta\right)/ (\varepsilon x)^2\right] \right\} \ . 
\end{equation}
Integration over $\theta$ gives
\begin{equation} \label{2e13b}
<\nu(\varepsilon)> \simeq \frac{1}{\sqrt{2\pi}}
\int^{\sqrt{2}/\varepsilon}_{-\sqrt{2}/\varepsilon} dy \, e^{-y^2/2} +
\frac{1}{\pi \sqrt{3}}  \int^1_0 dx \int^\pi_0 d\varphi \, \exp \left[ -
\frac{1}{\varepsilon^2} \left(\cos ^2 \varphi +  \frac{3 \sin ^2 \varphi}{x^2}
\right)\right] 
\ . \end{equation} 
To understand these results, consider first the limit $\varepsilon \to 0$. In
this case, the coefficients $\left\{ a_k\right\}$ tend to have a very narrow
distribution centered around zero, and the polynomial (\ref{2e7}) is well
approximated by 
$$ P(z) \buildrel{\varepsilon \to 0}\over{\simeq }1 + z^N  \ ; $$ 
the roots of this polynomial are given by Eq.(\ref{1e3}), with $r_N = 1$ and
$\phi=\pi/N$. This is indeed the behaviour recovered from Eqs.(\ref{2e13a}) and
(\ref{2e13b}). In (\ref{2e13b}), when $\varepsilon \to 0$ the first term in the
r.h.s. tends to one while the second tends to zero, implying $<\nu> = 1$.
Moreover, in (\ref{2e13a}) $\exp [-\cos ^2(\frac{N}{2} \theta)/\varepsilon^2]
\to 0 $ for any $\theta$ except at $\theta= \frac{2\pi}{N} (k+ \frac{1}{2})$,
$k= 0, \ldots, N-1$ where we get a delta peak, in agreement with the crystal
distribution. 

On the other extreme, when $ \varepsilon \to \infty$, we can neglect in
(\ref{2e7}) the term $1+z^N$  and 
\begin{equation} \label{2e14p}
P (z) \buildrel{\varepsilon \to \infty}\over{\simeq}  \sum^{N-1}_{k=1} a_k
\, z^k \ , \;\;\;\;\;\;\;\;\;\;\;\;\; a_{N-k}= \bar a_k \ . \end{equation} 
In this limit, the first term in the r.h.s. of Eq.(\ref{2e13b}) tends to zero,
while the double integral of the second term tends to $\pi$, and therefore 
\begin{equation} \label{2e14}
<\nu (\varepsilon)> \buildrel{\varepsilon \to \infty}\over{\simeq} 1/\sqrt{3},
\end{equation}
while from (\ref{2e13a}) we recover a uniform density   
\begin{equation} \label{2e15}
<\rho (\theta)> \buildrel{\varepsilon \to \infty}\over{\simeq
}\frac{N}{2\pi \sqrt{3}}. 
\end{equation} 
Eq.(\ref{2e14}) answers the question of how efficient is the SI-symmetry to
locate roots of a random polynomial on ${\cal C}$. We observe that indeed it
has a strong effect in the distribution of roots, since it manages to locate a
fraction $1/\sqrt{3} \simeq 57\%$ of the roots exactly on the unit circle. (For
a different proof of this result see Ref.~\cite{dun}.) 

Fig.1b shows the superposition of the roots of $200$ iterations of a $N=48$
self-inversive polynomial with all the second moments equal. The total number
of zeros is the same as in Fig.1a; the strong concentration of roots on $\cal
C$ is stressed in that figure by the reduction of the black intensity outside
$\cal C$. 

Fig.2 displays the fraction of points lying on  $\cal C$  as a function of
$\varepsilon$, Eq.(\ref{2e13b}). For small $\epsilon$ we observe the existence
of a plateau which can be interpreted in the following way. At $\epsilon =0$,
as we said before the roots coincide with the crystal lattice and $<\nu>=1$.
When $\epsilon$ increases, and since the zeros are analytic functions of that
parameter, they cannot immediately move outside $\cal C$ because this would
violate the self-inversive symmetry (zeros come by symmetric pairs with respect
to ${\cal C}$). The only way they can get out from ${\cal C}$ is to first move
 along ${\cal C}$
until two roots become degenerate, and then split in the radial direction, one
zero moving in the positive radial direction, the other towards the origin. The
size of the plateau can be estimated as the typical perturbation needed to
produce a coalescence of two roots starting from the crystal solution. 

It is instructive to compare our results to an analogous result due to M.
Kac \cite{kac}. He considers the case of polynomials with {\sl real}
coefficients $\left\{ a_k\right\}$ having a GRI distribution with all the second
moments equal. These polynomials satisfy the functional equation 
$$\overline{P(z)} =P(\bar z)$$
(the roots lie either on the real axis or come by symmetric pairs under
reflection with respect to it). He computes the fraction of real roots and
finds a much smaller effect of the symmetry as compared to Eq.(\ref{2e14}),
since he shows that $<\nu> \sim \ln N/N$ as $N\to \infty$ (he also considers the
distribution of the zeros on the real axis, see \cite{kac}). A numerical
simulation of the distribution of roots of random polynomials with real
coefficients is included in Fig.1c. The weak concentration of roots on the real
axis can be appreciated in the figure from the fact that the density of points
surrounding $\cal C$ is essentially unchanged as compared to Fig.1a. For
completeness, we plot in Fig.1d the distribution of roots of 200 trials of
$N=48$ SI-polynomials with real coefficients. See also subsection 4.2 for an
analogous result concerning SU(2) polynomials. 

\vspace{0.2in}

To conclude this subsection, let us point out that there is a simple
way to put {\sl all} the roots of a self-inversive polynomial on the unit
circle $\cal C$ (in a statistical sense). In appendix A we prove that the 
general formula for the fraction of roots lying on ${\cal C}$ for SI polynomials
of the form
\begin{equation} \label{2e15b}
P (z) = \sum^N_{k=0} a_k \, z^k \ , \;\;\;\;\;\;\;\;\;\;\;\;\; a_{N-k}= 
\bar a_k \end{equation} 
where the $\left\{ a_k \right\}$ are complex GRI distributed variables is given
by 
\begin{equation} \label{2e18}
<\nu>= \frac{2}{N} \sqrt{\frac{g_2}{g_1} } \end{equation}
where
\begin{equation} \label{2e19}
g_1= \sum^{\frac{N-1}{2}}_{k=1} \sigma^2_k \quad, \;\;\;\;\;\;\;\;\; g_2=
\sum^{\frac{N-1}{2}}_{k=1}\left(\frac{N}{2}-k\right)^2 \sigma^2_k  \ .
\end{equation} 
This is an exact formula valid for arbitrary $N$ and arbitrary variances 
$\sigma_k^2$ of the coefficients. The particular case of equal variances is
explicitly written in Eq.(\ref{2e31a}) below. This result was also obtained
recently in \cite{ek}, where the reader can also find a geometrical 
interpretation of it. Consider now the parametrization 
$$ \sigma_k = k^s \ , $$
where $s$ is an arbitrary real number. Using the asymptotic expansion
$$\sum^L_{k=1} k^\alpha \simeq \frac{L^{\alpha +1}}{\alpha +1} +O(L^\alpha),
\quad \;\;\;\;\;\;\;\; \alpha > -1 \ , $$ 
from Eqs.(\ref{2e19}) and (\ref{2e18}) it follows, to leading order in $N$ 
\begin{equation} \label{2e20} 
<\nu> \simeq \frac{1}{ \sqrt{(s+1)(2s+3)}}, \quad \;\;\;\;\;\;\;\; 
s > - \frac{1}{2} \ .\end{equation}
For $s=0$ (all the $\sigma$'s equal), we recover the previous result 
$<\nu> =1/\sqrt{3}$. For $s \to -1/2$, then $<\nu> \to 1$. For $s=-1/2$ we can 
estimate the rate of convergence towards $<\nu> =1$ as $N \to \infty$ from 
\begin{equation} \label{2e21}
\sum^L_{k=1} k^{-1} \simeq \ln N + C+ O(1/N) \end{equation}
(where $C$ is Euler's constant). Then, from (\ref{2e18}) and (\ref{2e19}) we 
get
\begin{equation} \label{2e22}
<\nu> \simeq 1-\frac{3}{4} \, \frac{1}{\ln (N/2)}+ O(1/N),  \quad 
\;\;\;\;\;\;\;\;\;\; s=-1/2 \ . 
\end{equation}
The convergence is therefore quite slow (for example, to put 98\% of the zeros 
on $\cal C$ we need $N \simeq 3.8 \times 10^{16}$). 

\subsection{Correlations between the roots}

Having determined that the SI symmetry locates, in the large-$N$ limit,
$1/\sqrt{3}$ of the zeros on $\cal C$, the next relevant question is what are 
the correlations existing between those roots. In particular, we would like to 
know if they repel each other or not and how their correlations compare to 
the random matrix theory (RMT). We have therefore computed, using the same
techniques as before, the average two-point correlation function 
$$
R_2(\tau)= <\rho(\theta) \rho(\theta +\tau)>$$
for the set of roots lying on $\cal C$  for polynomials of the form
(\ref{2e15b}). For this kind of polynomials the function $R_2(\tau)$ does not
depend on $\theta$ since the distribution of roots is invariant under
rotations.

The exact result valid for arbitrary $N$ and arbitrary variances is
presented in appendix B (cf Eq.(\ref{b15})). In the particular case of 
equal variances and in the limit $N \rightarrow \infty$, $\tau\to 0$, $N\tau\to
{\rm constant}$ the two-point correlation function normalized to the square of
the mean density (\ref{2e15}) takes the form 
\begin{equation} \label{2e16}
{\tilde R}_2 (\delta) = R_2 (\delta)/(N/2 \pi \sqrt{3})^2 = 
\frac{24}{\sqrt{C}} 
\left[B \arcsin \left(\frac{B}{A}\right)+\sqrt{A^2-B^2} \right], 
\end{equation}
where
$\delta=\tau N/2\pi$ and
\begin{equation} \label{2e17} \begin{array}{lll}
A=&\frac{1}{8(\pi \delta)^2} \left\{ \frac{(\pi \delta)^2}{3} -\left[\cos(\pi
\delta)-\frac{\sin (\pi \delta)}{\pi \delta}\right]^2 / \left[ 1-\frac{\sin
^2(\pi \delta)}{(\pi \delta)^2} \right] \right\} \\ 
B=&\frac{1}{4(\pi \delta)^2} \left\{ \cos (\pi \delta) + \left[ \frac{(\pi
\delta)^2}{2}-1 \right] \frac{\sin(\pi \delta)}{\pi \delta} - \frac{1}{2}
\frac{\sin(\pi \delta)}{\pi \delta} \left[\cos(\pi \delta)-\frac{\sin (\pi
\delta)}{\pi \delta}\right]^2 / \left[1-\frac{\sin ^2(\pi \delta)}{(\pi
\delta)^2} \right] \right\} \\ 
C=& 1 - \frac{\sin^2 (\pi \delta)}{(\pi \delta)^2} \ .\end{array}
\end{equation} 

Fig.3 is a plot of the function ${\tilde R}_2 (\delta)$. For short distances
there is a repulsion between the zeros. More precisely, from (\ref{2e16}) it
follows that 
$${\tilde R}_2 (\delta)  \buildrel{\delta \to 0}\over{\simeq} \frac{\pi^2}{10 
\sqrt{3}} \, \delta \ . $$
This behaviour is reminiscent of the linear repulsion between eigenvalues
obtained for the orthogonal ensemble (GOE) of the random matrix theory, but
with a different slope. However our computations correspond to systems without
time reversal invariance, since the coefficients are complex. We therefore
could have expected a quadratic repulsion, like in the GUE case. This
particular point deserves further investigation. The long-range behavior shows
on the other hand pronounced oscillations. 

We have also computed  numerically the nearest-neighbour spacing distribution
for the zeros lying on $\cal C$, shown  in Fig.4. This distribution was
investigated in Ref.~\cite{long}, where a recursive procedure for $p(\delta)$ 
was developed. Our numerical results are in agreement with those obtained in 
that reference.

From the general results of Appendix B, it is also possible to compute the
two-point correlation function for the case $\sigma_k =k^{-1/2}$, when
statistically $100\%$ of the zeros lie on ${\cal C}$. We now find that the
correlations between zeros tend to be much closer to a crystal-like solution
(i.e., strong oscillations that survive for large values of $\delta$) than in
the case of constant second moments. Fig.5 displays the function ${\tilde 
R}_2 (\delta)$ computed analytically from Eq.(\ref{b15}) for a self-inversive
polynomial with $\sigma_k=k^{-1/2}$ and $N=3601$; for such a value of $N$ the
fraction of roots lying on $\cal C$ is, from Eq.(\ref{2e22}), $<\nu>=0.9$. In
the light of the results of section 2, the emergence of a crystal-like
behaviour when $\sigma_k = k^{-1/2}$ is to be expected since in this case $N
\eta_k \to 0$ as $N\rightarrow\infty$. 

\subsection{The variance of the number of zeros lying on ${\cal C}$}

In section 3.1 we have established that SI polynomials of the form
(\ref{2e15b}) have asymptotically on the average a fraction $<\nu>=1/\sqrt{3}$
of their roots lying on ${\cal C}$ if the complex coefficients $\{a_k\}$ are
GRI-distributed with the same second moment. Our purpose now is to compute the
asymptotic behaviour of the variance of that average fraction, defined by 
\begin{equation} \label{2e31}
\sigma_\nu^2=<\nu^2> - <\nu>^2 .
\end{equation}
From Eqs.(\ref{2e18}) and (\ref{b16}) the exact form of the average number of 
roots lying on ${\cal C}$ for arbitrary $N$ and equal variances is
\begin{equation} \label{2e31a}
<\nu> = \sqrt{\frac{1/3+1/N+2/(3N^2)}{1+1/N}} \simeq 
\frac{1}{\sqrt{3}} \left(1+\frac{1}{N}\right) + O(1/N^2)
\end{equation}
and therefore
\begin{equation} \label{2e32}
<\nu>^2 \simeq \frac{1}{3} + \frac{2}{3 N}.
\end{equation}

On the other hand,
\begin{equation} \label{2e33} \begin{array}{ll}
<\nu^2> & = \frac{1}{N^2} \int_0^{2\pi}\int_0^{2\pi}<\rho(\theta)\rho(\theta')>
             d\theta \, d\theta' \nonumber \\
        & = \frac{1}{N^2} \left[\int_0^{2\pi}<\rho(\theta)> d\theta +
            \int_0^{2\pi}\int_0^{2\pi}R_2(\theta,\theta') d\theta \, d\theta' 
            \right] \nonumber \\
        & \simeq \frac{1}{N^2}\left[ \frac{N}{\sqrt{3}}+2 \int_0^{2\pi} (2\pi-
          \tau) \, R_2(\tau)  d\tau \right] + O(1/N^2) \end{array} 
\end{equation} 
where we have replaced the average number of roots on ${\cal C}$ by its
asymptotic value $N/\sqrt{3}$ and we have also exploited the fact that
$R_2(\theta, \theta')$ depends only on the difference $\tau$ between $\theta$ and
$\theta'$ to express the double integral as a simple one. Because $R_2(\tau)$ 
is symmetric with respect to $\tau=\pi$, Eq.(\ref{2e33}) can be rewritten as 
$$
<\nu^2>\simeq\frac{1}{\sqrt{3}N}+ \frac{4\pi}{N^2}\int_0^{\pi} R_2(\tau) \,
d\tau 
$$
or, again normalizing $R_2(\tau)$ to the square of the asymptotic mean density 
${\tilde R}_2 (\tau)=R_2 (\tau)/(N/2 \pi \sqrt{3})^2$
\begin{equation} \label{2e34}
<\nu^2>\simeq\frac{1}{\sqrt{3}N}+ \frac{1}{3\pi}\int_0^{\pi}
{\tilde R}_2 (\tau)\,d\tau. 
\end{equation}
As shown in Fig.3, in the large-$N$ limit ${\tilde R}_2 (\tau)$ has some
oscillations on a scale $\tau \sim O(1/N)$ and tends to one for larger values
of $\tau$. In order to separate the contribution to the integral in
Eq.(\ref{2e34}) from the oscillatory part of ${\tilde R}_2 (\tau)$, we rewrite 
it in the form 
\begin{equation} \label{2e35}
\int_0^{\pi} {\tilde R}_2 (\tau)\,d\tau=\frac{2\pi}{N}\int_0^{q} {\tilde R}_2 
(\delta,N)\,d\delta + \int_{2\pi q/N}^{\pi} {\tilde R}_2 (\tau,N)\,d\tau \ ,
\end{equation}
where $\delta=\tau N/2\pi$ and $q$ is a parameter which is large compare to one
but much smaller than $N$. In Eq.(\ref{2e35}) we have explicitly indicated the
$N$-dependence of $R_2$. Asymptotically (cf appendix B) 
$$ \begin{array}{ll}
{\tilde R}_2 (\delta,N) & \rightarrow {\tilde R}_2 (\delta) + O(1/N)\\
{\tilde R}_2 (\tau,N) & \rightarrow 1 + 2/N + O(1/N^2) \end{array}
$$
where ${\tilde R}_2 (\delta)$ is given by Eq.(\ref{2e16}). Therefore, keeping 
terms up to order $1/N$
\begin{equation} \label{2e36}
\int_0^{\pi} {\tilde R}_2 (\tau)\,d\tau \simeq \frac{2\pi}{N}\int_0^{q} {\tilde 
R}_2 (\delta)\,d\delta + \pi -\frac{2\pi q}{N}+\frac{2\pi}{N} =
\pi +\frac{2\pi }{N} \left( 1- \int_0^{q} \left[1-{\tilde R}_2 (\delta)\right] 
\, d\delta \right). 
\end{equation}
Because ${\tilde R}_2 (\delta)$ tends to one as $\delta\to\infty$, we are
allowed to take the limit $q\to\infty$ in the integral. 

Collecting Eqs.(\ref{2e36}), (\ref{2e34}), (\ref{2e32}) and (\ref{2e31})
we finally get
\begin{equation} \label{2e37}
\sigma_\nu^2 \simeq \left( \frac{1}{\sqrt{3}} - \frac{2}{3} \Delta\right)\frac{1}{N},
\end{equation}
with $\Delta=\int_0^\infty\left[1-{\tilde R}_2 (\delta)\right]\, d\delta$. We
were not able to compute analytically this integral, and made instead a
numerical calculation. We obtained $\Delta\simeq 0.44733$.

\section{$SU(2)$ Polynomials}
\setcounter{equation}{0}

\subsection{A theorem concerning the ergodicity of wave-functions}

In the previous section we were mainly concerned with the problem of locating
on $\cal C$ as many as possible of the roots of a  random polynomial and
studying their correlations. This problem concerns in particular the spectral 
statistics of asymptotic approximations of chaotic systems. In this section we 
explore a somewhat different and in some sense opposite problem: polynomials
whose roots distribute uniformly (at least as $N \to \infty$) on certain
surfaces.  Our motivation is related to the asymptotic $ \hbar \to 0$ structure
of quantum eigenstates of classically chaotic systems. 

Consider a system characterized by its total angular momentum $\vec J= (J_x,
J_y, J_z)$ whose modulus $J$ is conserved by the dynamics. The motion of the
arrow $\vec J$ in the three-dimensional space can be represented by a point
moving on the surface of a two-dimensional sphere, a Riemann sphere denoted 
${\cal S}$, which is in fact the phase space of the system.

The equations of motion for $\vec J$ are such that the representative point is
assumed to move on ${\cal S}$ in a chaotic way. This is possible if the angular
momentum $\vec J$ is coupled to some external time-dependent field, typically
a magnetic field. The simplest case is a periodic time-dependence, which we
henceforth assume. Integrating the equations of motion of $\vec J$ over one
period of the field, the classical dynamics for the point moving on the surface
of the sphere reduces to a discrete map $M$ acting on ${\cal S}$ 
\begin{equation} \label{3e1}
{\vec J}^{ (n+1)} = M \left( {\vec J}^{ (n)} \right) .
\end{equation}
These equations determine the position of the arrow at time $t=n+1$ knowing its
position at time $t=n$. 

The quantization of such a map introduces a one-period evolution operator $U$, 
the analog of the classical map $M$
\begin{equation} \label{3e2}
|\psi^{(n+1)}> = U |\psi^{(n)}> \ , 
\end{equation}
where $|\psi^{(n)}>$ defines the quantum state of the system at time $t=n$.
Because the modulus of $\vec J$ is conserved, then $[U , {\vec J}^{ 2}]=0$ and
the Hilbert space is finite and $(2J+1)$-dimensional. We can choose as a basis
of that space the eigenstates of $J_z$, $J_z|m>= \hbar m|m>$, $ m=-J, -J+1,
\ldots , J$. In particular, the eigenstates of the unitary operator $U$ 
\begin{equation} \label{3e3}
U |\psi_\alpha> = {\rm e}^{{\rm i} \omega_\alpha} |\psi_\alpha>,
\;\;\;\;\;\;\;\;\;\;\; \alpha=1,\ldots,2J+1 
\end{equation}
can be written as
$$|\psi_\alpha> = \sum^J_{m=-J} a^{(\alpha)}_m |m> \ .$$

The classical limit of such models corresponds to $N=2J \to \infty$. For
convenience we normalize the radius of the sphere, given by $\hbar
\sqrt{J(J+1)}$, to one. 

In order to have a unified semiclassical framework for both classical and
quantum mechanics, it is convenient to introduce a phase-space representation
of the Hilbert space. For that purpose, we project the eigenstates $|\psi_
\alpha>$ into $SU(2)$ coherent-states $|z>$ \cite{klau,per},
$\psi_\alpha (z)= <z|\psi_\alpha>$ with the result (we drop from now on the 
subscript $\alpha$)
\begin{equation} \label{3e4}
\psi (z) = \sum^N_{k=0} \sqrt{ C^k_N} \, a_k \, z^k ,
\end{equation} 
where $N= 2J$, $C^k_N$ are the binomial coefficients and where we have shifted 
to the new label $k=  m+J$. 

The complex variable $z$ labeling the coherent states and appearing in the
polynomial (\ref{3e4}) is connected to the variables $(\theta, \phi)$ spanning
the Riemann sphere by a stereographic projection of the plane onto the sphere,
$z=\cot\left(\theta/2\right) {\rm e}^{{\rm i} \phi}$. The function
$\psi (z) $ is therefore an analytic function defined on the
two-dimensional sphere. Because it is a polynomial of degree $N$, it has $N$
zeros in that space which completely determine (up to a global normalizing
factor) the quantum state. 

Our purpose here is to analyze the structure of the eigenstates of $U$ in a
regime where the classical dynamics is dominated by chaotic trajectories; this
means that classically the iteration of a typical initial point covers in time
the entire two-dimensional sphere ${\cal S}$ in a more or less uniform way.
According to the correspondence principle, in the semiclassical limit the
quantum states of such a system must tend, at least in a weak sense, to the
microcanonical density \cite{voros}. The simplest asymptotic realization would
be a function $\psi (z)$ whose modulus is uniform over ${\cal S}$. This,
however, is not an allowed solution since $\psi (z)$ has to have $N$
zeros in ${\cal S}$, and therefore cannot be uniform. Moreover, the number of
zeros proliferates in the semiclassical limit. The closest approximation to a
uniform density would then be a function $\psi (z)$ whose zeros spread
all over ${\cal S}$. This behaviour was in fact already observed in
Ref.\cite{lv} for eigenstates of chaotic systems. In the following, we make a
precise statement concerning the ergodicity of the distribution of zeros for
such systems. Assuming that the coefficients $\left\{ a_k \right\}$
in Eq.(\ref{3e4}) are GRI-distributed, we prove that the zeros of $\psi (z)$ 
are indeed spread all over ${\cal S}$ and that their distribution is moreover
{\sl uniform}. 

Our assumption concerning the coefficients $\left\{ a_k \right\}$ is
motivated by the random matrix theory. As is well known, the statistical
properties of the spectrum of classically chaotic systems are well described,
in the universal regime, by the results of the RMT \cite{bgs,mehta}. Much less
explored are the eigenstates of such systems, {\sl i.e.} the statistical
properties of the coefficients $\left\{ a_k \right\}$ for chaotic
systems and how they compare to the RMT (some results concerning this problem
can be found in \cite{fh}). The invariance under unitary transformations of the 
GUE ensemble of random matrices implies that the joint distribution function 
for the amplitudes must be 
$$
{\cal D}_{RMT} ({\vec a}) = \frac{1}{| S_{2(N+1)} |} \delta \left[ 1 - 
\sum^N_{k=0} |a_k |^2 \right]
$$
where $|S_n| = 2 \pi^{n/2}/\Gamma (n/2)$ is the surface of a 
$(n-1)$-dimensional sphere of unit radius. When computing average properties of 
the zeros of $\psi (z)$, this distribution is strictly equivalent to a 
GRI distribution (see ref. \cite{kac}, p.6)
\begin{equation} \label{3e5} 
{\cal D} (a_k) = \frac{1}{(2 \pi)^N} \exp \left\{ -\frac{1}{2} \sum^N_{k=0}  
\left| a_k \right|^2 \right\} \ .
\end{equation} 
Thus, the amplitudes $\left\{ a_k \right\}$ in RMT turn out to be gaussian 
uncorrelated random variables.

Assuming this distribution function, our purpose now is to compute the
associated average density of zeros on ${\cal S}$, $<\rho (\theta, \varphi)>$,
and compare it to the "ergodic" distribution conjectured above. Expressed in
terms of the complex variable $z$, the density of zeros can be written (see
appendix C) 
\begin{equation} \label{3e6}
\rho (z)= \delta [\psi (z)] \left| \frac{d \psi}{d z}\right|^2. 
\end{equation}
The next step in the computation would then be to exponentiate both terms in the
r.h.s. of Eq.(\ref{3e6}) and compute the ensemble average over the
coefficients. We find, however, that it is not necessary to exponentiate the
Jacobian $\left| d \psi / d z \right|^2$. From Eqs.(\ref{3e4}) and
(\ref{3e6}), using the exponential expression of the delta function and
computing the ensemble average we find, for {\sl arbitrary $N$}, the result 
\begin{equation} \label{3e7} 
<\rho (z)> d^2 z=\frac{N}{\pi} \frac{d^2z}{(1+|z|^2)^2} = \frac{N}{4\pi} \sin
\theta \,  d\theta d\phi \end{equation} 
i.e., a uniform  distribution of the zeros on the Riemann sphere. The proof of
Eq.(\ref{3e7}) is given in appendix C. 

The result (\ref{3e7}) constitutes a precise statement concerning the
ergodicity of the (zeros of ) eigenstates of chaotic systems, and makes a
connection between the concept of ergodicity and the RMT. Since the zeros are 
genuine wavefunction parameters, their equidistribution constitutes a stronger 
statement than just the limiting ergodicity of the Husimi function (which is a 
smooth quantity and bilinear in $\psi$). 

As already mentioned, Eq.(\ref{3e7}) holds for arbitrary $N$. In  particular
in the extreme case $N=1$ ($J=1/2$) it means that the single root $z_1 =
-a_0/a_1$ of the monomial $\psi(z) =a_0+a_1 z$ is uniformly distributed on the
Riemann sphere if $a_1$ and $a_0$ are complex variables having a GRI
distribution and the same second moment. This result can be directly checked
from Eq.(\ref{1e4}), 
$$ {\cal D} (a_0, a_1) d^2 a_0 d^2 a_1 =\frac{1}{(2\pi)^2 \sigma^4}  \exp
\left\{ -\frac{1}{2\sigma^2} \left[ |a_0|^2 +|a_1|^2 \right] \right\} d^2 a_0
d^2 a_1 , $$ 
$$ \Rightarrow \;\; {\cal D} (z_1, a_1) d^2 z_1 d^2 a_1 =\frac{1}{(2\pi)^2
\sigma^4}  \exp \left\{ -\frac{|a_1|^2}{2 \sigma^2} \left[ 1+|z_1|^2 \right]
\right\} |a_1|^2 d^2 z_1 d^2 a_1 \ . $$ 
Integrating over $a_1$ we find 
$$ {\cal D} (z_1) =<\rho(z_1)>= \frac{1}{\pi}
\frac{1}{\left(1+|z_1|^2\right)^2} $$ 
in agreement with (\ref{3e7}). A polynomial like (\ref{3e4}) generalizes this
result to  arbitrary $N$. The validity of the equidistribution of roots for
arbitrary $N$ is however related to the gaussian nature of the distribution of
the coefficients. We have numerically tested other distributions. For example,
we find that for a uniform distribution of the coefficients in the interval
[-1,1], the distribution of roots tends to be uniform only for large values of
$N$. 

To conclude this subsection, let us mention that correlations between roots of 
random SU(2) polynomials were recently studied in \cite{ls} and compare to 
results obtained from quantum chaotic systems. Moreover, the general $k$-point 
correlation functions were also recently computed analytically by J. Hannay 
\cite{han}.

\subsection{A physical example} 

In order to illustrate the theorem (\ref{3e7}) we consider a kicked-spin model, 
classically defined by the Hamiltonian 
$$ H=\frac{p}{2}
J^2_z + \mu J_x \sum^\infty_{n=-\infty} \delta (t-n) $$ 
where $\mu$ and $p$ are constant parameters. Integrating over a period $\Delta 
t=1$ the equations of motion take the form of a discrete map 
\begin{equation} \label{3e8} 
\vec J^{ (n+1)} = R_x(\mu) \, R_z \left(pJ_z^{(n)}\right) \vec J^{ (n)}.
\end{equation} 
$R_i(\lambda)$ represents a rotation around the $i$-th axis by an angle
$\lambda$, and $\vec J= (J_x, J_y, J_z)$. The quantum mechanical analog of
Eq.(\ref{3e8}) is the one-step unitary  operator 
\begin{equation} \label{3e9} 
U={\rm e}^{-\frac{{\rm i}}{\hbar} \mu J_x} {\rm e}^{-\frac{{\rm i}}{\hbar}
\frac{p}{2} J^2_z} \end{equation} 
acting on a $(2J+1)$-dimensional Hilbert space. The stationary equation 
(\ref{3e3}) determines the eigenphases $\omega_\alpha$  and eigenstates 
$|\psi_\alpha>$ of $U$. According to the result Eq.(\ref{3e7}), for parameters 
$(\mu,p)$  for which the classical map (\ref{3e8}) is dominated by chaotic 
trajectories we expect the zeros of the polynomials (\ref{3e4}) associated with
the eigenstates of $U$ to be uniformly distributed over the Riemann sphere.  
Before checking this, let us briefly mentioned some properties of $U$.

The operator $U$ is not generic since it has two symmetries: it 
commutes with two antiunitary operators \cite{hks}
$$T_1= e^{i\pi J_z} {\rm e}^{{\rm i}\mu J_x} K,  \quad \quad T_2= {\rm
e}^{-{\rm i}\mu J_x} {\rm e}^{{\rm i}\pi J_y} K, $$ 
where $K$ is the usual antiunitary complex conjugation operator. These 
operators satisfy $T^2_1= T^2_2=1$ and the time reversal property $T_1 U T_1= 
T_2 U T_2=U^{-1}$. These two symmetries are nongeneric in the sense that (a) 
they are not just the  conjugation operator usually connected to time-reversal
invariance and (b) they depend  on the parameter $\mu$ controlling, together
with $p$, the dynamics of the  system. 

Because $T^2_i=1$, the existence of an operator $T_i$ commuting with $U$
implies that  the eigenstates of $U$ can always be chosen $T$-invariant,
$T|\psi_\alpha>= |\psi_\alpha>$. In the coherent-state representation, this
latter equation imposes  a functional equation on each polynomial
$\psi_\alpha(z)$ 
\begin{equation} \label{3e10}
\overline{\psi_\alpha (\tau_i (z))} = \psi_\alpha(z) \quad \quad i=1, 2\ ;
\;\;\;\;\;\; \alpha=1, \ldots ,2J+1\ , \end{equation}
where $\tau_i , \, i=1,2$ are  the classical versions of the quantum operators
$T_i$ which associates to each point of the classical phase space $z$ an image
$\tau_i (z)$. Each of these transformations has a symmetry line, defined by the
set of points $z$ satisfying $\tau_i (z)=z$. In terms of the canonical
conjugate variables ($\phi, \cos \theta)$ spanning the sphere, the two symmetry
lines are given by 
\begin{equation} \label{3e11}
\cos \theta =\pm \frac{\sin \phi}{ \sqrt{\sin ^2\phi + \frac{1\mp \cos \mu}
{1\pm \cos \mu} } } \ , 
\end{equation}
where the upper and lower sign holds for the $T_1$ and $T_2$ symmetry, 
respectively.

Because of (\ref{3e10}), if $z_k$ is a root of $\psi (z)$, then $\tau_i
(z_k)$ will also  be a root. According to the results of section 3, if the
roots are symmetric with respect to a line and if the coefficients of the
polynomial are random, we expect a concentration of roots over that line. In
the present case, we have two symmetry lines. Figure 6a shows the
superposition of the 60 roots of the 61 eigenstates obtained by a numerical
diagonalization of (\ref{3e9}) for $J =30$, $\mu =1$ and $p=4\pi$, which
classically looks fully chaotic \cite{leb1}. We  observe the expected
concentration of roots over the two  symmetry lines, a free-of-roots region 
close to them, and a tendency to cover in a more or less uniform way the
remaining phase space (see ref.\cite{ls} for more details). Although we haven't
computed analytically, we suspect that the concentration of roots over those
lines is not macroscopic (i.e., asymptotically tends to zero), like the number
of real roots of random polynomials having real coefficients. In this context,
let us mention that it has been shown recently \cite{ek} that the asymptotic
fraction of real roots of an SU(2) polynomial having real coefficients is
$1/\sqrt{N}$. This should be compare to the $\log N / N$ fraction valid for the
original problem proposed by Kac. 

In order to break both antiunitary symmetries $T_1$ and $T_2$ we add an extra
term in the propagator 
\begin{equation} \label{3e12}
U = {\rm e}^{-\frac{{\rm i} t}{2 \hbar} J_z^2} {\rm e}^{-\frac{{\rm i}}{\hbar}
\mu J_x} {\rm e}^{-\frac{{\rm i}}{2 \hbar}p J_z^2} . 
\end{equation}
Figure 6b and 6c show the superposition of the roots of the $61$ eigenstates
for the same $p$ and $\mu$ as in Fig.6a for $t =1$ and $t=6$, respectively.
After a transition regime where the symmetry lines are still observed (even
though the symmetry has been broken), for $t=6$ the roots spread in a more or
less uniform way over the whole phase space, as predicted in Eq.(\ref{3e7}). 

Notice that although in order to have better statistics we have superimposed in
the figures the roots of all the eigenstates, the ergodic theorem (\ref{3e7})
holds for {\sl individual} eigenstates. However, a direct numerical test based
on a single eigenstate would need much higher values of $J$. 

For other systems having different phase spaces, it is natural to expect -- 
guided by semiclassical intuition -- a result equivalent to Eq.(\ref{3e7}), at
least in the semiclassical limit. And indeed a related result was recently 
found \cite{leb2} for the usual Bargmann representation \cite{barg} of quantum 
mechanics. This is a representation of Hilbert space in terms of entire
functions 
\begin{equation} \label{ce1}
\psi (z) = \sum^{\infty}_{k=0} \, \frac{a_k}{\sqrt{k!}} \, z^k .
\end{equation}
The associated phase space is the two-dimensional plane labelled by the 
canonical variables $(q,p)$, and $z=(q-{\rm i} p)/\sqrt{2}$. It was shown 
that if the coefficients $\left\{ a_k \right\}$ in Eq.(\ref{ce1}) 
are GRI-distributed with all the second moments equal, the zeros of $\psi (z)$
are uniformly distributed over the whole plane. Moreover, if the sum in
Eq.(\ref{ce1}) is truncated at a finite value $N$, then the density of roots 
is uniform inside a circle of radius $\sqrt{N}$, and tends to zero outside.
Generalizations to other geometries as well as the autocorrelation function of 
$\psi (z)$ have also been considered \cite{leb2}.

\section{Concluding remarks}

We have studied the distribution and correlation of roots of random polynomials
under several conditions. In section 3, we have explored, motivated by the
problem of the semiclassical spectral properties of chaotic systems, different
ways to increase the number of roots of a random polynomial located on ${\cal 
C}$, and studied correlations between the roots. If all the coefficients of the
characteristic polynomial have the same standard deviation, then asymptotically
the self-inversive symmetry locates a fraction $1/\sqrt{3}$ of the roots on
$\cal C$ with a variance inversely proportional to the degree of the polynomial
(section 3.3). The two-point correlation function of these zeros behaves
linearly for short distances (Fig.3). However, if the standard deviations are
not equal for all the coefficients but instead are given by $\sigma_k=
k^{-1/2}$, then the fraction of roots lying on ${\cal C}$ tends to one as $N
\to \infty$. But the convergence is slow (logarithmic), and the two-point
correlation function much more crystalline-like (Fig.5). 
Surprisingly, and unlike the case of standard polynomials with real
coefficients (the problem of Kac), we were able to compute explicitly  
the {\sl exact} fraction of roots lying on ${\cal C}$ and their two-point
correlation function for arbitrary $N$ and arbitrary variances. 

To improve by less artificial means the number of roots lying on ${\cal C}$ 
we need to incorporate additional correlations between the coefficients. However
no simple procedure exists, and the exact conditions for all the roots of the
characteristic polynomial to lie on $\cal C$ are certain complicated
determinantal inequalities for the coefficients (see e.g. \cite{marden}). 

The self-inversive property arises naturally in other problems of physics, as in
the case of certain Ising models in statistical mechanics where the partition
function takes a polynomial form when written in terms of the fugacity
\cite{yl}. There, the SI-symmetry is connected to a spin-up spin-down
(or particle-hole) symmetry of the system. As shown in \cite{yl}, under certain
assumptions all the necessary correlations implying unitarity are present and
as a consequence all the roots of the partition function  lie on $\cal C$. In
Ref.~\cite{derrida} the reader can find additional examples and references
concerning the distribution of roots of partition functions in connection with
the theory of phase transitions. 

Another  famous example of concentration of the roots of a certain function on
some simple  curve in the complex  plane is provided by the Riemann
zeta-function $\zeta (z)$. This function, which has no explicit random
parameters entering its definition, satisfies the functional equation $\xi
(1-z) =\xi (z)$, where  $\xi (z)= \pi^{-z/2} \Gamma (z/2) \zeta (z)$.
Accordingly, its roots are symmetric with respect to the critical line $\Re e
(z) =1/2$. The Riemann hypothesis asserts that all the nontrivial roots of 
$\zeta (z)$ lie on that line.

In section 4 we have shown that if we assume for the coefficients $\left\{ a_k
\right\}$ a GRI distribution with second moments equal to $\sqrt{C_{N}^{k}}$
then the roots of the eigenstates of a classically chaotic spin system,
Eq.(\ref{3e4}), are uniformly distributed over the phase space (in this case,
the two-dimensional sphere).

\newpage

\renewcommand{\appendix}
       {\par
        \setcounter{section}{0}
        \setcounter{subsection}{0}
        \gdef\afterthesectionpunctdefault{:}
        \gdef\thesection{{Appendix \Alph{section}}}
        \gdef\theequation{A\arabic{equation}}
        \gdef\lefteqno{\Alph{section}}
        \setcounter{equation}{0}}

\appendix
\section{}
\setcounter{equation}{0}
\renewcommand{\theequation}{\Alph{section}.\arabic{equation}} 

In this appendix we compute the average density and average fraction of
roots lying on ${\cal C}$ for a self-inversive polynomial of the 
form 
\begin{equation} \label{a1}
P(z)=1+ \sum^{N-1}_{k=1} a_k z^k+z^N \ , \;\;\;\;\;\;\; a_{N-k} = {\bar a}_k
\end{equation} 
for arbitrary $N$. The coefficients $a_k$ are assumed to be complex independent 
variables having a Gaussian distribution
\begin{equation} \label{a2} 
{\cal D}(a_1, \ldots ,a_M)= \frac{1}{(2\pi)^M \prod \sigma_k^2} \exp \left(
-\frac{1}{2} \sum^M_{k=1} |a_k|^2/\sigma^2_k \right) \ ,
\end{equation}
where $M=(N-1)/2$ ($N$ is an odd integer). By the substitution $z=\exp ({\rm
i}\theta)$, the problem reduces to the computation of the average density and
average number of roots of a real function in the interval $0\leq \theta <2\pi$
(cf. Eq.(\ref{2e9})) 
\begin{equation} \label{a3}
f(\theta)= \frac{1}{2}{\rm e}^{-{\rm i} N \theta/2} P(\exp ({\rm i}\theta))
=\cos\left(\frac{N} {2} \theta\right) + \sum^M_{k=1} \left\{c_k \cos
\left[\left(\frac{N}{2} -k\right)\theta\right]+ d_k \sin
\left[\left(\frac{N}{2}-k\right) \theta\right]\right\} 
\end{equation}
where $c_k=\Re e (a_k) \ \; d_k =\Im m (a_k)$. The derivative of this function
with respect to $\theta$ is 
\begin{equation} \label{a4}
f'(\theta)=-\frac{N}{2} \sin (\frac{N}{2} \theta) -\sum^M_{k=1}
\left\{\left(\frac{N}{2} -k\right) c_k \sin
\left[\left(\frac{N}{2}-k\right)\theta\right]-\left(\frac{N}{2}-k\right) d_k
\cos \left[\left(\frac{N}{2} -k\right)\theta \right] \right\}\ . 
\end{equation}
The density of roots of $f(\theta)$ is defined by 
$$
\rho (\theta)=\delta[f (\theta)] \, |f'(\theta)| 
$$
and using the Kac's representation (\ref{2e11}) for $\delta[f]$
and $|f'|$  we get
\begin{equation} \label{a5}
\rho(\theta)= \frac{1}{2\pi^2}\int^\infty_{-\infty} d\xi \, {\rm e}^{{\rm i}
\xi f(\theta)} \int^{\infty}_{-\infty} d\eta \frac{ 1-{\rm e}^{{\rm i}\eta
f'(\theta) }}{\eta^2} \ . 
\end{equation} 
In order to compute the average of $\rho(\theta)$ over the ensemble (\ref{a2}),
we proceed in the following way. We first replace $f(\theta)$ and $f'( \theta)$
in (\ref{a5}) by its definition Eqs.(\ref{a3}) and (\ref{a4}). Then we average
$\rho(\theta)$ over the coefficients $c_k$ and $d_k$ (it can be shown that the
order of integration can be interchanged; see [kac]),
\begin{equation} \label{a6}
 <\rho(\theta)> =\int^{\infty}_{-\infty}  \ldots \int^{\infty}_{-\infty} 
 d^2 a_1 \ldots d^2 a_{M} \rho (\theta) {\cal D} (a_1, \ldots , a_{M}) \ . 
\end{equation}
The final step is to integrate over $\eta$ and $\xi$.

\noindent {\bf 1.} We will need the average of 
$e^{i \xi f} e^{i\eta f'}$. From (\ref{a3}) and (\ref{a4}) we have
\begin{equation} \label{a7} 
\xi f+\eta f'  = \xi \cos \left(\frac{N}{2} \theta\right) -\frac{N}{2} \eta
\sin  \left(\frac{N}{2} \theta\right) + \sum^M_{k=1} (s_k c_k+t_kd_k) \ , 
\end{equation}
where
\begin{equation} \label{a8}
\begin{array}{l}
s_k=\xi \cos \left[\left(\frac{N}{2}-k\right)\theta\right]-(\frac{N}{2}-k) \eta
\sin \left[\left(\frac{N} {2}-k\right)\theta\right] \\ 
t_k= \xi\sin \left[\left(\frac{N}{2}-k\right)\theta\right]+(\frac{N}{2}-k) \eta
\cos \left[\left(\frac{N} {2}-k\right)\theta\right] \ . \end{array} 
\end{equation}
From these equations and the fact that
\begin{equation} \label{a9}
\frac{1}{\sqrt{2\pi \sigma^2_k}} \int^{+\infty}_{-\infty} 
{\rm e}^{-a_k^2/(2\sigma_k^2) +b_k a_k} \ da_k = {\rm e}^{b_k^2 \sigma_k^2/2} 
\end{equation}
we get from (\ref{a6})
$$ <{\rm e}^{{\rm i}\xi f} {\rm e}^{{\rm i}\eta f'}>= \exp \left\{ {\rm
i}\left[ \xi \cos\left(\frac{N}{2}\theta\right) -\frac{N}{2} \eta \sin
\left(\frac{N}{2} \theta\right)\right]\right\} \exp\left\{-\frac{1}{2}
\sum^M_{k=1} (s_k^2+t^2_k)\sigma_k^2\right\} \ . 
$$
But $s^2_k+t^2_k=\xi^2+(N/2-k)^2 \, \eta^2$. Then defining 
\begin{equation} \label{a10}
\begin{array}{l}
g_1= \sum\limits^M_{k=1} \sigma^2_k \\
g_2= \sum\limits^M_{k=1} \left(\frac{\displaystyle N}{\displaystyle 2}
-k\right)^2 \sigma^2_k \end{array} 
\end{equation}
we can write
\begin{equation} \label{a11}
<{\rm e}^{{\rm i}\xi f} {\rm e}^{{\rm i}\eta f'}>= \exp \left[-\frac{1}{2}
(g_1\xi^2+g_2 \eta^2)\right] \exp \left\{ {\rm i}[\cos \left(\frac{N}{2}
\theta\right)\, \xi-\frac{N}{2} \sin \left(\frac{ N}{2} \theta\right)\, \eta]
\right\} \ . 
\end{equation}
Note that the functions $g_1$ and $g_2$ contain all the information concerning 
the variances of the random coefficients.
For $\eta=0$, the result is
\begin{equation} \label{a12}
<{\rm e}^{{\rm i} \xi f}>=\exp \left(-\frac{1}{2} g_1 \xi^2\right) \exp
\left[{\rm i} \cos (\frac{N}{2} \theta)\, \xi\right] \ . 
\end{equation}
Then from (\ref{a11}), (\ref{a12}) and (\ref{a5})
\begin{equation} \label{a13} 
<\rho(\theta)>=\frac{1}{2\pi^2} \int^\infty_{-\infty} d\xi \, {\rm
e}^{-\frac{1}{2} g_1 \xi^2+i \cos (\frac{N}{2} \theta)\, \xi} 
\int^\infty_{-\infty} \frac{d\eta}{\eta^2} \left(1-{\rm e}^{-\frac{1}{2}
g_2\eta^2- {\rm i}\frac{N}{2} \sin(\frac{N}{2} \theta) \eta}\right) \ . 
\end{equation} 

\noindent {\bf 2.} The next step is to evaluate the integrals in (\ref{a13}).
The integral over $\eta$ can be written
$$I(\beta)= \int^{\infty}_{-\infty} \frac{d \eta}{ \eta^2} \left(1-{\rm
e}^{-\beta \eta^2} {\rm e}^{-{\rm i} \frac{N}{2} \sin (\frac{N}{2} \theta)
\eta} \right) 
$$
where $\beta=g_2/2$. Since
$$ \frac{\partial I}{\partial \beta}=\int d\eta \, {\rm e}^{-\beta \eta^2} {\rm
e}^{-{\rm i} \frac{N}{2} \sin (\frac{N}{2} \theta)\eta} =
\sqrt{\frac{\pi}{\beta}}\, {\rm e}^{-\frac{N^2}{16} \sin^2(\frac{N}{2}
\theta)/\beta} 
$$
and, from (\ref{2e11})
$$I(\beta=0)=\int^\infty_{-\infty} \frac{d \eta}{\eta^2}\left(1-{\rm e}^{-{\rm
i} \frac{N}{2} \sin (\frac{N}{2} \theta)\, \eta}\right)=\frac{\pi N}{2} \left|
\sin \left(\frac{N}{2} \theta\right)\right| \ , 
$$
then
\begin{equation} \label{a14}
I(\beta)=\frac{\pi N}{2} \left| \sin \left(\frac{N}{2}
\theta\right)\right|+\sqrt{\pi} \int^\beta_0 \frac{d y}{\sqrt{y}} {\rm
e}^{-\frac{N}{16}  \sin ^2(\frac{N}{2} \theta)/y } \ . 
\end{equation}
Moreover, the integral  over $\xi$ in (\ref{a13}) is 
\begin{equation} \label{a15}
\int^\infty_{-\infty} d\xi {\rm e}^{-\frac{1}{2} g_1\xi^2 +{\rm i}\cos
(\frac{N}{2} \theta)\xi}=\sqrt{\frac{2\pi}{g_1}} {\rm e}^{-\cos ^2 (\frac{N}{2}
\theta ) / 2g_1} \ . \end{equation} 
Doing the change of variable $x=\sqrt{y/\beta}$ in the integral of
Eq.(\ref{a14}), the final result for the average density of zeros on ${\cal C}$
is 
\begin{equation} \label{a16}
<\rho(\theta)>=\frac{{\rm e}^{-\cos ^2(\frac{N}{2} \theta)/2g_1}}{\sqrt{2\pi
g_1}} \left\{\frac{N}{2}\left| \sin \left(\frac{N}{2} \theta\right)\right| +
2\sqrt{\frac{g_2}{2\pi} } \int^1_0 dx \, {\rm e}^{-N^2 \sin^2 (\frac{N}{2}
\theta)/(8g_2 x^2)} \right\} \ . 
\end{equation}

\noindent {\bf 3.} The average fraction of zeros lying on ${\cal C}$ is defined
as 
\begin{equation} \label{a17}
<\nu>=\frac{1}{N} \int^{2\pi}_0 <\rho(\theta)>d\theta \ . \end{equation}
The integral involving the first term between curly brackets in (\ref{a16})
can be rewritten (by an obvious change of variables)
$$ \int^{2\pi}_0 {\rm e}^{-\cos^2 (\frac{N}{2} \theta) / (2g_1)  } \left| \sin
\left(\frac{N}{2} \theta\right)\right| d\theta = \frac{2}{N} \int^{N\pi}_0 d
\varphi \, {\rm e}^{-\cos ^2 (\varphi)/(2g_1) } |\sin \varphi| \ . 
$$
The function to be integrated is periodic of period $\pi$, and since $\sin 
\varphi$ is a positive function in that interval the previous equation takes 
the form
\begin{equation} \label{a18}
2\int^\pi_0 d\varphi \, {\rm e}^{- \frac{1}{2g_1} \cos^2 \varphi } \sin \varphi
= 2 \sqrt{g_1} \int^{1/\sqrt{g_1}}_{-1/\sqrt{g_1}}d y {\rm e}^{-y^2/2} \ . 
\end{equation}
The integral over the second term between curly brackets in (\ref{a16}) can be 
written, putting $\varphi = N \theta/2$ and by the same argument as before
\begin{equation} \label{a19} 
\int^1_0 dx  \int^{2\pi}_0 d\theta \, {\rm e}^{-\left[\cos ^2 (\frac{N}{2}
\theta)/g_1 +N^2 \sin ^2 (\frac{N}{2} \theta)/(4g_2 x^2)\right]/2} =
2\int^1_0dx \int^\pi_0 d\varphi \, \exp \left\{-\left[ \frac{cos^2
\varphi}{g_1}+ \frac{N^2}{4g_2} \frac{\sin^2 \varphi}{x^2} \right]/2\right\}\ . 
\end{equation}
Putting together Eqs.(\ref{a16})-(\ref{a19}), the fraction of zeros lying on 
${\cal C}$ is given by
\begin{equation} \label{a20}
<\nu>= \frac{1}{\sqrt{2\pi}} \int^{1/\sqrt{g_1}}_{-1/\sqrt{g_1}}
 d y \, {\rm e}^{-y^2/2} +\frac{2}{\pi N} \sqrt{ \frac{g_2}{g_1} }
\int^1_0 dx \int^\pi_0 d\varphi \, {\rm e}^{-\left(
\frac{1}{2g_1} \cos ^2 \varphi + \frac{N}{8 g_2} \frac{\sin ^2
\varphi}{x^2}\right)} \ . 
\end{equation} 
Both  results (\ref{a16}) and (\ref{a20}) are valid for arbitrary $N$ and 
arbitrary variances $\sigma^2_k , k=1, \ldots , M$. Moreover, both expressions
are the sum of two terms: the first one in each expression is related
to the term  $1+z^N$ in (\ref{a1}), while the second comes from the random 
part  of the polynomial. In the case where all the variances are equal,
$\sigma^2_k =\sigma^2 \ \forall k,$ then the coefficients $g_1$ and $g_2$
reduce to
\begin{equation} \label{a21} \begin{array}{l}
g_1=\sigma^2 \sum\limits^M_{k=1} 1= \sigma^2 (N-1)/2\\
g_2= \sigma^2 \sum\limits^M_{k=1} (\frac{\displaystyle N}{\displaystyle 2}-k)^2
=\sigma^2 \left( \frac{\displaystyle N^3}{\displaystyle 24} -
\frac{\displaystyle N^2}{\displaystyle 8}+\frac{\displaystyle N}{\displaystyle
12} \right) \ . \end{array} \end{equation} 
In the limit $N\to \infty$, $g_1 \simeq \sigma^2 N/2=\epsilon^2/2$,
$g_2\simeq \sigma^2N^3/24= \epsilon^2N^2/24$, where we have introduced 
the scaled parameter $\epsilon =
\sigma \sqrt{N}$. Replacing these asymptotic expressions for $g_1$ and $g_2$ in
(\ref{a16}) and (\ref{a20}), we find that $<\rho (\theta)>$ and $<\nu>$ depend
only on the rescaled parameter $\epsilon$, and are given by Eqs.(\ref{2e13a})
and (\ref{2e13b}), respectively.

On the other hand, when $|\sigma_k|>>1$ $\forall k$, then $g_1$ and $g_2 \to
\infty$ and we can ignore the first term in (\ref{a20}). In this case (the
pure random limit, which corresponds to neglect the term $z^N+1$ in $P(z)$),
the fraction of roots lying on ${\cal C}$, Eq.(\ref{a20}), reduces to
Eqs.(\ref{2e18}) which in turn simplifies to Eq.(\ref{2e14}) if all the 
$\sigma$'s are equal and if we keep only the leading term in $1/N$. Moreover, 
the average density of zeros tends to 
\begin{equation} \label{a22} 
<\rho(\theta)> \to \frac{1}{\pi} \sqrt{\frac{g_2}{g_1}} \ . 
\end{equation} 

\section{}

We compute here the average two-point correlation function
\setcounter{equation}{0}
\begin{equation} \label{b1} 
R_2 (\tau)= <\rho(\theta)\rho(\theta+\tau)>
\end{equation}
for the roots lying on ${\cal C}$ of a self-inverse random polynomial of the
form 
\begin{equation} \label{b2}
P (z)= \sum^N_{k=0} a_k \, z^k \ , \;\;\;\;\;\;\;\; a_{N-k} ={\bar a}_k
\end{equation} 
where the coefficients $a_k \ , k=0 , \ldots \ , \frac{N-1}{2} =M$ are complex
independent variables having a Gaussian distribution (\ref{a2}) (we again
assume for simplicity that $N$ is odd; for even $N$, we must take $M=N/2$). 
Substituting $z= \exp (i \theta)$ in (\ref{b2}) and ignoring prefactors we end 
up with the real function 
$$f(\theta)=\sum^M_{k=0} \left\{ c_k \cos
\left[\left(\frac{N}{2}-k\right)\theta\right]+d_k \sin \left[
\left(\frac{N}{2}-k\right)\theta\right]\right\} \eqno(B.3a)$$ 
where $c_k=\Re e (a_k)$, $d_k=\Im m (a_k)$, and whose derivative is
$$f'(\theta)=-\sum^M_{k=0}\left\{ (\frac{N}{2}-k) c_k \sin
\left[\left(\frac{N}{2}-k\right) \theta\right]-\left(\frac{N}{2}-k\right) d_k
\cos \left[\left(\frac{N}{2}-k\right)\theta\right]\right\} \ . 
\eqno(B.3b)$$
The function $f(\theta)$ has, as shown in appendix A, an average number 
$N/\sqrt{3}$ of zeros in the interval $0\leq \theta<2\pi$. Moreover, since
$<\rho(\theta)>$ is in this case independent  of $\theta$ (cf. Eqs.(\ref{a22})
and (\ref{2e15})), $R_2$ depends on $\tau$ but not on $\theta$. Using the 
definition (\ref{2e10}) of  $\rho(\theta)$ and Eqs.(\ref{2e11}), we can write
the two-point correlation function as
\setcounter{equation}{3}
\begin{equation} \label{b4}
\rho(\theta)\rho(\theta+\tau)=\frac{1}{4\pi^2}  \int\!\int\!\int\!\int^\infty_{
-\infty} d\xi_1 d\xi_2 \frac{d\eta_1}{\eta^2_1} \frac{d \eta_2}{\eta^2_2}
 {\rm e}^{{\rm i}\xi_1 f(\theta)} {\rm e}^{{\rm i}\xi_2 f(\theta+\tau)}
\left(1- {\rm e}^{{\rm i}\eta_1 f'(\theta)}\right)\left(1-{\rm e}^{{\rm
i}\eta_2 f'(\theta+\tau)}\right) \ . 
\end{equation}

We proceed as in appendix A. We first replace Eqs.(B.3) in 
(\ref{b4}) and compute the average over the coefficients $(c_k, d_k)$ according
to the definition (\ref{a6}). Thenceforth we  evaluate the integrals over
$\xi_i$ and $\eta_i$. 

\noindent {\bf 1.} We will need $<{\rm e}^{{\rm i} [\xi_if(\theta)+\xi_2
f(\theta+\tau)+\eta_1 f'(\theta) +\eta_2 f'(\theta+\tau)] }>$.
From Eqs.(B.3) 
$$ \xi_1f(\theta)+\xi_2f(\theta+\tau) +
\eta_1f'(\theta)+\eta_2f'(\theta+\tau)  =\sum^M_{k=0} [(s_k+u_k)c_k+
(t_k+v_k)d_k ]
$$
where
\begin{equation} \label{b5}
\begin{array}{l}
{\displaystyle s_k= \xi_1 \cos \left[\left(N/2-k\right)
\theta\right]-\left(N/2-k\right)\eta_1 \sin \left[\left( N/2 - k\right) \theta
\right] } \\ 
{\displaystyle t_k= \xi_1 \sin \left[\left(N/2-k\right)
\theta\right]+(N/2-k)\eta_1 \cos \left[\left( N/2 -k\right) \theta \right] } \\
{\displaystyle u_k= \xi_2 \cos \left[\left(N/2-k\right)
(\theta+\tau)\right]-\left(N/2-k\right)\eta_2 \sin \left[\left( N/2-k\right)
(\theta+\tau)\right] } \\ 
{\displaystyle v_k= \xi_2 \sin \left[\left(N/2-k\right) (\theta+\tau)\right]
+\left(N/2-k\right)\eta_2 \cos \left[\left(N/2-k\right) (\theta+\tau) \right] \
. } 
\end{array} \end{equation}
Averaging over the Gaussian ensemble gives
$$<{\rm e}^{{\rm i}[\xi_1f(\theta)+\xi_2f(\theta+\tau)+\eta_1f'(\theta)+\eta_2
f'(\theta+\tau)]}>= \exp \left\{-\frac{1}{2} \sum^M_{k=0}[(s_k+u_k)^2+(t_k
+v_k)^2] \sigma^2_k\right\} \ . $$ 
But
$$ \begin{array}{ll} (s_k+u_k)^2+(t_k+v_k)^2 = &\xi^2_1+\xi^2_2 +(N/2-k)^2
(\eta^2_1 +\eta^2_2) \\ 
&+2\left[\xi_1\xi_2+(N/2-k)^2\eta_1\eta_2\right]\cos
\left[\left(N/2-k\right)\tau\right] \\ 
& + 2(N/2-k)(\xi_2\eta_1- \xi_1 \eta_2) \sin \left[\left(N/2-k\right)\tau\right]
\ , \end{array} $$
which is independent of $\theta$, as it should be. Then defining
\begin{equation} \label{b6}
\left\{ \begin{array}{l}
g_1= \sum\limits^M_{k=0} \sigma^2_k \\
g_2= \sum\limits^M_{k=0}(N/2-k)^2\sigma^2_k \\
g_3= \sum\limits^M_{k=0} \cos[(N/2-k)\tau]\sigma^2_k \\
g_4= -\partial g_3/ \partial \tau = \sum\limits^M_{k=0}(N/2-k)\sin 
     [(N/2-k) \tau]\sigma^2_k\\
g_5= -\partial^2 g_3/ \partial \tau^2 = \sum\limits^M_{k=0}(N/2-k)^2 
     \cos [(N/2-k) \tau]\sigma^2_k \end{array} \right. 
\end{equation}
we obtain the result
\begin{equation} \label{b7} \begin{array}{l} 
<\exp \left\{{\rm i}[\xi_1f(\theta)+\xi_2f(\theta+\tau)
+\eta_1f'(\theta)+\eta_2f'(\theta+ \tau) ]\right\} > = \\ 
\;\;\;\;\;\;\;\;\;\;\;\;\;\;\;\;\; \exp\left\{
-[g_1(\xi^2_1+\xi_2^2)+g_2(\eta^2_1 +\eta^2_2)+2(g_3\xi_1\xi_2 +g_5\eta_1
 \eta_2)+2g_4(\xi_2\eta_1-\xi_1\eta_2)]/2\right\} \ . 
\end{array} \end{equation}
From this expression the average of the different terms appearing in
(\ref{b4}) can be evaluated, with the result
\begin{equation} \label{b8} \begin{array}{ll}
R_2(\tau) =&\frac{1}{4\pi^4} \int\!\int\!\int\!\int^\infty_{-\infty} d \xi_1
d\xi_2 \frac{\displaystyle d\eta_1}{\displaystyle \eta^2_1} \frac{\displaystyle
d \eta_2}{\displaystyle \eta^2_2} e^{ -[g_1(\xi^2_1+\xi^2_2)+2g_3
\xi_1\xi_2]/2} \times \\ 
&\left\{ 1-e^{-(g_2\eta^2_1+2g_4 \xi_2\eta_1)/2} - 
e^{-(g_2\eta^2_2-2g_4 \xi_1 \eta_2)/2} + e^{-
[g_2(\eta^2_1+\eta^2_2)+2 g_5 \eta_1\eta_2+2g_4 (\xi_2 \eta_1-\xi_1\eta_2)
]/2} \right\} . 
\end{array} \end{equation}

\noindent {\bf 2.} The next step is to compute the integrals in (\ref{b8}).
The integrals over $\xi_1$ and $\xi_2$ are straightforward, since
they involve exponentials of quadratic forms:
\begin{equation} \label{b9}
R_2(\tau)=\frac{1}{2\pi^3 \sqrt{g^2_1 -g^2_3}}\int\!\int^\infty_{-\infty}
\frac{d\eta_1}{\eta^2_1}\frac{d \eta_2}{\eta^2_2} \left[1-{\rm
e}^{-\frac{\alpha}{2} \eta^2_1}-{\rm e}^{-\frac{\alpha}{2} \eta^2_2} + {\rm
e}^{-\frac{\alpha}{2}(\eta^2_1+\eta^2_2)} {\rm e}^{-\beta \eta_1 \eta_2}
\right] \ , 
\end{equation}
where we have introduced
\begin{equation} \label{b10}
\alpha=g_2-\frac{g_1\,g_4^2}{(g_1^2-g^2_3)} 
 \ , \;\;\;\;\;\;\;\;\;\;\; \beta = g_5 - \frac{g_3\,g_4^2}{(g^2_1-g^2_3)} \ . 
\end{equation}
Consider now the integral
\begin{equation} \label{b11} 
I(\beta)=\int\int^\infty_{-\infty}  \frac{d\eta_1}{\eta_1^2}
 \frac{d \eta_2}{\eta^2_2} \left[1-{\rm e}^{-\frac{\alpha}{2}\eta^2_1 }
 - {\rm e}^{-\frac{\alpha}{2}\eta^2_2 } +{\rm e}^{-\frac{\alpha}{2}(\eta^2_1+
\eta^2_2)}  {\rm e}^{-\beta \eta_1 \eta_2} \right] \end{equation} 
and write it in the form
\begin{equation} \label{b12}
I(\beta)=\left. I(\beta)\right|_{\beta=0} +\left. \frac{\partial I}{\partial
\beta} \right|_{\beta=0} \, \beta+ \int\int^\beta_0 dy I(y) \ . \end{equation} 
From Eq.(\ref{a15}) evaluated at $\theta=0$ we get the first term of the 
previous equation
\begin{equation} \label{b13}
\left. I(\beta)\right|_{\beta=0} = \int\int^\infty_{-\infty} \frac{d \eta_1} {
\eta_1^2} \frac{d\eta_2}{\eta^2_2} \left(1-e^{-\frac{\alpha}{2}
\eta^2_1}\right) \left(1-e^{-\frac{\alpha}{2} \eta^2_2}\right)=2\pi \alpha \ . 
\end{equation}
Moreover
$$
\frac{\partial I}{\partial \beta}=-\int\int^\infty_{-\infty}
\frac{d\eta_1}{\eta_1} \frac{d\eta_2}{\eta_2} {\rm e}^{-\frac{\alpha}{2}
(\eta^2_1+\eta^2_2)} {\rm e}^{-\beta \eta_1\eta_2} \ ,
$$
implying $\left. \partial I/\partial \beta\right|_{\beta=0}=0$ by antisymmetry.
Furthermore,
$$ \frac{\partial^2 I}{\partial \beta^2}= \int \int^\infty_{-\infty} d\eta_1
d\eta_2 e^{-\frac{\alpha}{2}(\eta^2_1+\eta^2_2)}e^{-\beta \eta_1\eta_2}
= \frac{2\pi}{\sqrt{\alpha^2-\beta^2}} \ ,
$$
and integrating twice
\begin{equation} \label{b14}
\int^\beta_0dx \int^x_0 dy \frac{2\pi}{\sqrt{\alpha^2-y^2}}= 2\pi\left[\beta 
\arcsin (\beta/\alpha)+\sqrt{\alpha^2-\beta^2}-\alpha\right] \ . 
\end{equation}
We thus obtain, substituting in (\ref{b12}) the different terms
$$I(\beta)=2\pi \left[ \beta \arcsin \left(\frac{\beta}{\alpha}\right)+\sqrt{\alpha^2
-\beta^2}\right] \ .
$$
Then the exact average two-point correlation function, valid for arbitrary $N$
and arbitrary variances, is
\begin{equation}\label{b15} 
R_2(\tau)=\frac{1}{\pi^2 \sqrt{g^2_1-g^2_3}} \left[\beta \arcsin (\beta/\alpha)
+ \sqrt{\alpha^2-\beta^2}\right] \ . \end{equation}
The sums 
(\ref{b6}) -- needed in the computation of the coefficients $\alpha$ and 
$\beta$ -- can be evaluated explicitly if the variances are equal. We find
\begin{equation} \label{b16}
\left\{ \begin{array}{l}
g_1= \sigma^2 (N+1)/2 \\
g_2= \sigma^2 \left(N^3/24 + N^2/8 + N/12\right) \\
g_3= \sigma^2 \sin \left[ \tau(N+1)/2 \right]/\left[2\sin (\tau/2)\right] \\
g_4= -\frac{\sigma^2}{4\sin(\tau/2)} \left\{ N\cos\left[\tau(N+1)/2\right] -
      \frac{\sin(\tau N/2)}{\sin(\tau/2)}\right\} \\ 
g_5= \frac{\sigma^2}{4}\left\{\frac{\sin(\tau N/2)\cos(\tau /2)}{\sin(\tau /2)} 
     \left[\frac{N^2}{2} - \frac{1}{\sin^2(\tau /2)}\right] +N \cos(\tau N/2)
     \left[\frac{N}{2} + \frac{1}{\sin^2(\tau /2)}\right] \right\} \end{array}
     \right. 
\end{equation}

In this latter case, taking the limit $N\to\infty$, $\tau\to 0$, $N\tau\to 
{\rm constant}$, and normalizing $R_2$ to the square of the asymptotic mean 
density (\ref{2e15}), Eqs.(\ref{b15}), (\ref{b16}) and 
(\ref{b10}) can be rewritten in the form (\ref{2e16})-(\ref{2e17}). 

\section{}
We prove here the theorem Eq.(\ref{3e7}), stating that the average density
 of roots of a polynomial of the form
\setcounter{equation}{0}
\begin{equation} \label{c1}
\psi (z)= \sum^N_{k=0} \sqrt{ C^k_N} \, a_k \, z^k \  \end{equation}
is {\sl uniform} over the Riemann sphere for arbitrary $N$. In Eq.(\ref{c1})
the coefficients $a_k,\,  k= 0 \ , \ldots , N$ are assumed to be complex 
independent variables having a gaussian distribution with the same standard
deviation $\sigma_k=\sigma$ $\forall k$, and the $C^k_N$ are the binomial 
coefficients $C^k_N=N!/k!(N-k)!$.

We want to compute  the density of zeros of a complex function in the complex
plane. By definition
\begin{equation} \label{c2}
\rho(z)=\delta [\Re e \left( \psi (z)\right)]\,\delta[\Im m \left( \psi
(z)\right) ] \left| \begin{array}{ll} \frac{\partial \Re e (\psi)}
{\partial \Re e (z)} &\frac{\partial \Re e
(\psi)}{\partial \Im m (z)}\\ \frac{\partial \Im m (\psi)}
{\partial \Re e (z)} & \frac{\partial \Im m (\psi)}{\partial \Im m (z)}
\end{array} \right| 
\ . \end{equation}
By the Lemma 7.1, page 150 of Ref.\cite{rp} concerning the Jacobian of complex 
analytic functions, (\ref{c2}) can be rewritten
\begin{equation} \label{c3}
\rho(z)= \delta [\Re e \left( \psi (z)\right)]\, \delta[\Im m \left(
\psi(z)\right) ]\left| \frac{d \psi} {d z}\right|^2 \ . 
\end{equation}
We will use, for convenience, polar coordinates in the $z$-plane
$$ z= r \, {\rm e}^{{\rm i}\varphi} $$
and write $a_k=c_k+{\rm i} d_k$. Then, from (\ref{c1})
\begin{equation} \label{c4} \begin{array}{ll} 
f(r,\varphi)&= \psi (z=r\, {\rm e}^{{\rm i} \varphi}) \\
&=  \sum\limits^N_{k=0} \left\{ \sqrt{C_N^k} r^k [c_k \cos (k\varphi)-d_k \sin
(k \varphi) ] +{\rm i} \sqrt{ C_N^k} r^k [d_k \cos (k \varphi) +c_k \sin
(k\varphi) ] 
\right\} . \end{array} 
\end{equation}
Moreover
$$ \frac{d \psi}{d z} =\sum^N_{k=0} \sqrt{C_N^k} \, k \, a_k \, z^{k-1}
$$
and then, in polar coordinates
\begin{equation} \label{c5}
\left|\frac{d \psi}{d z}\right|^2= \sum^N_{k, \ell =0}
\sqrt{C_N^k C_N^\ell} (c_k +id_k)(c_\ell-id_\ell) r^{k+\ell-2}
{\rm e}^{{\rm i}(k-\ell)\varphi} k\, \ell \ . 
\end{equation}
To compute the average over the Gaussian ensemble
we will only exponentiate the delta functions in (\ref{c3}), but
not the Jacobian. Using the expression (\ref{2e11}) for the delta
functions, and (\ref{c4})-(\ref{c5}) for $f$ and the Jacobian, 
respectively, the density (\ref{c3}) can be expressed as
\begin{equation} \label{c6}
\begin{array}{ll}
\rho(r,\varphi) = &\frac{1}{(2\pi)^2} \int\!\int^\infty_{-\infty} d\xi_1
d\xi_2 \left\{ \sum\limits_{k=0}^N C_N^k k^2 (c^2_k+d^2_k)r^{2(k-1)} \right. \\
&\left. +\sum\limits^N_{k\neq \ell=0} \sqrt{C_N^k C_\ell^N} k\ell \left[c_k
c_\ell+d_k d_\ell+ {\rm i}(d_k c_\ell-c_k d_\ell)\right] r^{k+\ell-2}
{\rm e}^{{\rm i}(k-\ell)\varphi} \right\} \exp\left\{ \sum\limits^N_{n=0}
(\alpha_n c_n+\beta_n d_n)\right\} \end{array} 
\end{equation}
where
\begin{equation} \label{c7}
\begin{array}{l}
\alpha_n={\rm i} \sqrt{C_N^n}r^n [\cos (n \varphi)\xi_1 +\sin (n \varphi) \xi_2
] \\ 
\beta_n={\rm i} \sqrt{C_N^n}r^n [\cos (n \varphi)\xi_2 -\sin (n \varphi) \xi_1
] . 
\end{array} \end{equation}

\noindent {\bf 1.} The average over the coefficients $c_k$ and $d_k$ in 
(\ref{c6}) involves  expressions of the type 
$$ <c_k^{j_1} d^{j_2}_\ell \exp\left\{ \sum_{n=0}^N (\alpha_n c_n+\beta_n d_n)
\right\} > $$
where the symbol $<.>$ represents the average over the ensemble (\ref{1e4})
taking all variances $\sigma^2$ equal; the parameters $j_i$ in the latter
expression can take the values $1$ or $2$. The computation is straightforward;
for example, for $j_1=2$ and  $j_2=0$ 
$$ <c^2_k \exp\left\{\sum^N_{n=0} (\alpha_n c_n+\beta_n d_n)\right\}>
=\sigma^2(1+\alpha^2_k \sigma^2) \exp\left\{\frac{\sigma^2}{2} \sum^N_{n=0}
(\alpha^2_n+\beta^2_n)\right\} \ . 
$$
From (\ref{c7}) we can write
$$
\alpha^2_n +\beta^2_n= -C_N^n r^{2n} (\xi^2_1 + \xi^2_2) $$
and hence
$$ <c_k^2 \exp\left\{\sum_{n=0}^N (\alpha_n c_n+\beta_nd_n)\right\}>=\sigma^2
(1+\alpha^2_k \sigma^2) \exp\left\{-\frac{\sigma^2}{2}(1+r^2)^N
(\xi^2_1+\xi^2_1)\right\} \ . $$ 
The other averages are computed analogously. The result of averaging
(\ref{c6}) is 
\begin{equation} \label{c8} \begin{array}{ll}
<\rho(r,\varphi)> =& \frac{\sigma^2}{(2\pi)^2}
\int\int^\infty_{-\infty} d\xi_1d\xi_2\, \exp\left[-\frac{\sigma^2}{2}(1+r^2)^N
(\xi^2_1+\xi^2_2) \right] \left\{ \sum\limits^N_{k=0} C_N^k k^2
\left[2-\sigma^2 C_k^N r^{2k} (\xi^2_1+\xi^2_2) \right] \right.\\ 
&\left. +\sigma^2 \sum\limits^N_{k\neq\ell=0} \sqrt{C_N^k C_N^\ell} \, k \ell 
\, \left[\alpha_k \alpha_\ell+\beta_k\beta_\ell+{\rm
i}(\beta_k\alpha_\ell-\alpha_k\beta_\ell)\right] r^{k+\ell-2} {\rm e}^{{\rm
i}(k-\ell)\varphi} \right\}  \ . 
\end{array} \end{equation}

\noindent {\bf 2.} The integrals involving the first term between curly 
brackets in (\ref{c8}) give
\begin{equation} \label{c9} 
\int\int^\infty_{-\infty}d\xi_1d\xi_2 \left[2-\sigma^2 C_N^k
r^{2k}(\xi_1^2+\xi_2^2)\right] \exp\left\{-\sigma^2 (1+r^2)^N
(\xi_1^2+\xi_2^2)/2 \right\} = \frac{4\pi}{\sigma^2 (1+r^2)^N}
\left[1-\frac{C_N^k r^{2k}}{(1+r^2)^N} \right] \ . 
\end{equation}
Moreover, from Eqs.(\ref{c7}) it follows that
$$
\alpha_k\alpha_e+\beta_k\beta_\ell +{\rm i} (\beta_k\alpha_\ell- \alpha_k 
\beta_\ell )=-\sqrt{
 C_N^k C_N^\ell} r^{k+\ell} {\rm e}^{-{\rm i}(k-\ell)\varphi}
(\xi_1^2+\xi_2^2) \ . $$
Using this result, we evaluate the integrals involving the second term between 
curly brackets
\begin{equation} \label{c10} 
\int\int^\infty_{-\infty}d\xi_1d\xi_2 
{\rm e}^{-\sigma^2 (1+r^2)^N(\xi_1^2+\xi_2^2)/2} 
\left[\alpha_k\alpha_\ell+\beta_k\beta_\ell +i(\beta_k\alpha_\ell- \alpha_k 
\beta_\ell )\right] = - \frac{4\pi}{\sigma^4 (1+r^2)^N} \sqrt{  C_N^k
C_N^\ell}\, r^{k+\ell} {\rm e}^{-{\rm i}(k-\ell)\varphi} \ . 
\end{equation} 
Using Eqs.(\ref{c9}) and (\ref{c10}), (\ref{c8}) can be expressed as
\begin{equation} \label{c11} 
<\rho(r,\varphi)>=\frac{1}{\pi(1+r^2)^N} \left[\sum^N_{k=0} C_N^k k^2
r^{2(k-1)} -\frac{1}{(1+r^2)^N} \left(\sum^N_{k=0} C_N^k k\,r^{2k-1}\right)^2
\right] \ . 
\end{equation}
In order to evaluate the sums, we consider the identity
\begin{equation} \label{c12}
\sum^N_{k=0} C_k^N r^{2k}=(1+r^2)^N \ . \end{equation}
By differentiating Eq.(\ref{c12}) once and twice with respect to $r^2$ we get
\begin{equation} \label{c13}
\sum^N_{k=0} C_N^k k\, r^{2k-1}= r  \frac{\partial (1+r^2)^N}{\partial(r^2)}
= rN(1+r^2)^{N-1} \end{equation}
and
\begin{equation} \label{c14} \begin{array}{ll}
\sum\limits^N_{k=0} C_N^k k^2 \, r^{2(k-1)} &= r^2  \frac{\displaystyle
\partial^2 (1+r^2)^N}{\displaystyle \partial (r^2)^2}+\frac{\displaystyle
\partial(1+r^2)^N}{\displaystyle \partial (r^2)} \\ &=r^2N(N-1)(1+r^2)^{N-2}
+N(1+r^2)^{N-1} \ , \end{array} \end{equation} 
respectively.
Substitution of (\ref{c13}) and (\ref{c14}) into (\ref{c11}) gives, finally
\begin{equation} \label{c15}
<\rho(r, \varphi)> rdr d\varphi=\frac{N}{\pi} \frac{r}{(1+r^2)^2} dr d\varphi
= \frac{N}{\pi} \frac{d^2z}{(1+|z|^2)^2} \ . \end{equation}

A simpler form for the density of zeros is obtained projecting  (by a
stereographic projection from the north pole) the complex plane into the
two-dimensional Riemann sphere (having unit radius), spanned by the spherical
variables  ($\theta, \varphi$) 
$$ z= \cot \left(\theta/2\right) {\rm e}^{{\rm i} \varphi} \ . $$
This transformation explicitly shows that, in fact, the density (\ref{c15}) 
is {\sl uniform} on that surface
\begin{equation} \label{c16}
<\rho (\theta, \varphi)> d \theta d \varphi = \frac{N}{4\pi} \sin \theta \,
 d\theta d\varphi \ . 
\end{equation}

\pagebreak

\pagebreak

\large 
\begin{center} 
FIGURES
\end{center} 
\normalsize 

\begin{description}

\item{FIG. 1:} Distribution in the complex plane of the roots of random 
polynomials. In all the ((a) to (d)) cases we have superimposed the roots of
$200$ different trials of a polynomial of degree $N=48$ whose coefficients obey
a GRI distribution all having the same second moment $\sigma$. (a) Standard
random polynomial with all its coefficients complex and independent. (b) SI
random polynomial with complex coefficients; only half of the coefficients are
random, the other half being determined by complex conjugation. (c) standard
random polynomial with real coefficients. (d) SI random polynomial with real
coefficients. 

\item{FIG. 2:} The asymptotic average number of roots lying on the unit circle
${\cal C}$ for a SI random polynomial of the form (\ref{2e7}) as a function of
the parameter $\epsilon=\sigma\sqrt{N}$. 

\item{FIG. 3:} The asymptotic two-point correlation function ${\tilde R}_2$
(Eq.(\ref{2e16})) for the roots lying on ${\cal C}$ of SI polynomials with
complex GRI-distributed coefficients. 

\item{FIG. 4:} Nearest-neighbour spacing distribution for the case of Fig.3.

\item{FIG. 5:} The two-point correlation function ${\tilde R}_2$ for the roots
lying on ${\cal C}$ of $N=3601$ SI polynomials with complex GRI-distributed
coefficients and second moments $\sigma_k=k^{-1/2}$. 

\item{FIG. 6:} Phase-space distribution of the zeros of eigenstates of a
chaotic system. In the three parts of the figure we have superimpose the $60$
roots of the $N+1=61$ eigenstates obtained by numerical diagonalization of the
kicked-top map (\ref{3e12}) for $\mu = 1$ and $p= 4\pi$. (a) t=0, where two
antiunitary symmetries exist. We observe a concentration of roots over the two
associated phase-space symmetry lines given by Eq.(\ref{3e11}). (b) t=1, both
symmetries are now broken but there is still a concentration of roots on the
symmetry lines. (c) t=6, all vestiges of the symmetries have disappeared, and
we recover a uniform-like distribution, in agreement with the prediction
(\ref{3e7}). 

\end{description}

\end{document}